\documentclass[aps,a4paper,twocolumn]{revtex4-1}
\usepackage[pdftex]{graphicx}
\usepackage{float}
\usepackage{amssymb}   
\usepackage{epsfig}

\begin{document}

\title{Fixed-Node Diffusion Monte Carlo potential energy curve of the fluorine molecule F$_2$ using 
selected configuration interaction trial wavefunctions}

\author{Emmanuel Giner, Anthony Scemama, and Michel Caffarel}
\affiliation{Lab. Chimie et Physique Quantiques, CNRS-Universit\'e de Toulouse, France.
}

\begin{abstract}
The potential energy curve of the F$_2$ molecule is calculated with Fixed-Node Diffusion 
Monte Carlo (FN-DMC) using Configuration Interaction (CI)-type trial wavefunctions. 
To keep the number of determinants reasonable (the first and second derivatives of the trial wavefunction 
need to be calculated at each step of FN-DMC), the CI expansion is restricted to those determinants 
that contribute the most to the total energy. The selection of 
the determinants is made using the so-called CIPSI approach (Configuration Interaction using a Perturbative 
Selection made Iteratively). Quite remarkably, the nodes of CIPSI wavefunctions are found to be
systematically improved when increasing the number of selected determinants. To reduce the 
non-parallelism error of the potential energy curve a scheme based on the use of a $R$-dependent 
number of determinants is introduced. Numerical results show that improved FN-DMC energy 
curves for the F$_2$ molecule are obtained when employing CIPSI trial wavefunctions. 
Using the Dunning's cc-pVDZ basis set the FN-DMC energy curve is of a quality similar to 
that obtained with FCI/cc-pVQZ. A key advantage of using selected CI in FN-DMC 
is the possibility of improving nodes in a systematic and automatic way without resorting 
to a preliminary multi-parameter stochastic optimization of the trial wavefunction performed 
at the Variational Monte Carlo level as usually done in FN-DMC.
\end{abstract}


\maketitle



\newpage

\section {Introduction}
Fixed-Node Diffusion Monte Carlo (FN-DMC) and its diverse variants\cite{lester,rmp} are considered as accurate 
approaches for evaluating ground-state properties of molecules. Although it is certainly true for total energies, 
such a statement can be questioned when considering the (very) small energy differences involved in quantitative chemistry 
(atomization energies, energy variations along a chemical reaction path, forces viewed as infinitesimal energy differences, 
excitation energies, etc.), particularly for large systems. Although several sources of error make FN-DMC simulations 
non-exact, only the fixed-node error should be considered as truly fundamental, that is, conditioning
the practical limitations of the method. Other errors include the statistical error due to a finite number $N_{MC}$ of 
Monte Carlo steps, the time-step error resulting from the use of a finite time-step $\tau$ for propagating 
and branching walkers, and possible bias resulting either from 
a fluctuating finite population $M$ of walkers 
or a finite projection imaginary time $t$ when working at constant population size. 
However, in all cases such errors can be controlled and estimated through extrapolation techniques 
involving only {\it one single parameter} (that is, $N_{MC} \rightarrow +\infty$, $\tau \rightarrow 0$,
and $M$ or $t \rightarrow +\infty$). The error resulting from the use of a not-truly random number generator 
should also be added to this list but numerical experience has shown that its impact is generally (much) 
smaller than statistical fluctuations provided a sufficiently ``good'' pseudo-random generator is employed.\cite{lecuyer} 
In sharp contrast, the fixed-node error is much more 
challenging since, up to now, no simple and systematic scheme involving only a {\it finite} number of parameters 
exist for building up the exact $(3N-1)$-dimensional ($N$, number of electrons) nodal hypersurface that 
would suppress the fixed-node error. 
In FN-DMC the shape of the nodes are {\it implicitly} introduced via the trial wavefunction $\Psi_T$ 
used for propagating walkers. During the Monte Carlo simulation walkers are diffused (free Brownian motion) 
and moved deterministically using the drift vector, ${\bf \nabla} \Psi_T/\Psi_T$. Wherever $\Psi_T=0$ the drift diverges and 
walkers are pushed away from the nodal variety, thus imposing additional boundary conditions for the wavefunction 
(mathematically it means that the Schr\"odinger equation is solved stochastically with the additional condition that the solution vanishes wherever 
$\Psi_T$ vanishes). Common wisdom about nodes is that the more the salient 
physical/chemical features of the exact wavefunction are injected into the trial wavefunction the better nodes should be. 
Thus, an intense activity has been developed to introduce and test various functional forms for $\Psi_T$ taking into account 
important aspects of the wavefunction: Use of several more or less sophisticated forms for Jastrow-type prefactors describing 
local electron-electron (in particular, $r_{12} \rightarrow 0$ CUSP conditions) and electron-electron-nucleus interactions,\cite{mosko}
use of various multi-determinantal forms for introducing static (CASSCF-type wavefunctions)\cite{flad},\cite{fili}
or static/dynamic correlation effects (Valence Bond,\cite{braida,goddard,fili_vb} multi-Jastrow\cite{bouabca}, 
Configuration Interaction,\cite{qmcgrid}, etc.), use of geminal forms,\cite{mitas,sorella} backflow terms,\cite{rios} etc. 
Once the trial wavefunction form has been chosen a last step consists in optimizing stochastically the many parameters of $\Psi_T$ 
by minimizing the variational energy, its variance, or a combination of both.\cite{cyrus}
Depending on the system treated, this optimization may play an important role since the mixing of various terms of different origins may destroy 
the initial coherence of each wavefunction component and, thus, the nodal quality ({\it e.g.}, re-optimization of the Kohn-Sham or Hartree-Fock 
molecular orbitals in presence of the Jastrow term). A convenient feature of the fixed-node approximation is the variational property, 
$E_0^{FN} \geq E_0$ (equality for exact nodes), allowing a quantitative criterion for nodal quality.

Numerical experience for various molecular systems ranging from simple atoms and diatomics to bigger systems including hundreds 
of electrons has shown that fixed-node DMC total energies obtained from available trial wavefunctions are (very) accurate, the fixed-node error 
representing typically a small fraction of the total correlation energy (down to a few percents in the best cases).
Unfortunately, such a good precision is still insufficient to lead to reliable energy differences 
known to represent only a tiny fraction of the total correlation energy (typically, smaller or much smaller than 1$\%$). 
As a consequence, and similarly to all known {\it ab initio} approaches aiming at reaching chemical accuracy (including the most accurate 
highly-correlated wavefunction approaches) the quality of FN-DMC results for energy differences is tightly dependent on the level of 
error cancellation occurring when subtracting total energy components calculated separately (here, compensation of fixed-node errors).
Solving this problem is a major challenge faced by present FN-DMC approaches and motivates the search 
for better trial wavefunctions with better nodes and/or better procedures for optimizing and/or building accurate nodal hypersurfaces.

In this work, we present our first FN-DMC study of a full potential energy curve using our very recently proposed trial wavefunction\cite{canadian}
based on a perturbatively selected configuration interaction expansion (the so-called CIPSI algorithm\cite{cipsi1,cipsi2}).
Our first application to the ground-state energy of the oxygen atom using the truncated determinantal expansion generated by 
CIPSI has led to the lowest fixed-node energy so far (to the best of our knowledge) for this atom 
[99.4(1)$\%$ of the correlation energy recovered]. Applications to bigger systems are 
presently under way and appear to systematically lead 
to better nodes.\cite{preprint1,preprint2} An important feature of these applications is that 
the determinantal expansion built during the deterministic CIPSI selection process is used as it comes, no re-optimization of 
determinantal weights and molecular orbitals is performed. It is a particularly attractive feature since the usual many-parameter stochastic 
optimization step is thus avoided and to define a simple and automatic procedure (based on a purely deterministic algorithm) for optimizing nodes 
of arbitrary molecular systems is much simplified.
The aim of the present study is to investigate on the case of the F$_2$ molecule how the good results obtained on single-point calculations 
generalize or not when calculating potential energy curves. The various aspects conditioning the method: 
choice of the basis set, dependence on the number of determinants kept in the CI expansion, and coherence of the fixed-node error 
as a function of the internuclear distance (reducing the non-parallelism error) are investigated.

The contents of this paper is as follows. In Section \ref{qmc} a few words about the Fixed-Node Diffusion Monte Carlo method
employed here are given. In Section \ref{cipsi}, the CIPSI algorithm used for building the selected configuration interaction expansion
is presented. In Section \ref{results_f2} CIPSI and FN-DMC results for the F$_2$ ground-state potential energy curve and the 
corresponding spectroscopic quantities are presented in detail. The role played by the number of determinants selected and the coherence 
of the fixed-node error as a function of the internuclear distance are investigated.
Finally, the main ideas and results of this work are summarized in Sec. \ref{conclusion}.

\section{The Fixed-Node Diffusion Monte Carlo}
\label{qmc}
In this work the Fixed-Node Diffusion Monte Carlo (FN-DMC) method -the standard quantum Monte Carlo electronic-structure approach for molecules- is 
employed. For a detailed presentation of its theoretical and practical aspects, the reader is referred to the literature, 
{\it e.g} \cite{lester,towler,ency}. Here, we just recall that the central quantity of such approaches is the
trial wavefunction $\Psi_T$ determining both the magnitude of the fixed-node error through its approximate nodes as discussed in the 
introduction and the quality of the statistical convergence (good trial wavefunctions = small statistical fluctuations).
In the present case let us remark that the molecule is sufficiently small and the trial wavefunction sufficiently good to obtain statistical errors
much smaller than fixed-node ones. Thus, in practice, we will only be concerned with the problem of reducing as much as possible the fixed-node bias.
The computational cost of FN-DMC is almost entirely determined by the evaluation 
at each Monte Carlo step of the value of $\Psi_T$ and its first (drift vector) and second derivatives (Laplacian needed for the local energy).
In view of the very large number of MC steps usually required (typically at least millions and often much more) 
to be able of computing such quantities very rapidly is essential. 

In the present work, as presented in detail in the following section the trial wave function will be obtained from a truncated 
configuration interaction expansion, that is, a finite sum of determinants. Typically, the size of the expansion considered will range 
from a few thousands up to a few hundred thousands of determinants.
As a consequence, some care is required when computing such expansions to 
keep the computational cost reasonable.
The calculation of the Laplacian of the wave function and the drift term involves the computation of 
the inverse of the Slater matrices corresponding to each determinant.
At first glance, the CPU cost is expected to be proportional to the number of determinants $N_{\rm dets}$ involved in the expansion of the trial
wavefunction. Actually, it is not true since in the spin-free formalism used in QMC (Ref. \cite{matsen} and also \cite{lester,towler,ency}) each
Slater determinant expressed in terms of spin-orbitals decomposes into
a product of two determinants, each of them corresponding to a given occupation of a set
of {\it purely spatial} molecular orbitals.
In practice, only two inverse Slater matrices (one of each spin) are
computed with an ${\cal O}(N^3)$ algorithm.  All the other matrices are
built using ${\cal O}(N^2)$ Sherman-Morisson updates.\cite{sherman} Therefore,
the computational cost scales as ${\cal O}(N_{\alpha}^2 \times \sqrt{N_{\rm dets}})$ where $N_\alpha$ is the
number of $\alpha$ electrons and $N_{\rm dets}$ is the number of determinants (products of $\alpha$ and $\beta$ determinants).

\section{Perturbatively selected configuration interaction} 
\label{cipsi}
In multi-determinantal expansions the ground-state wavefunction $|\Psi_0\rangle$ 
is written as a linear combination of Slater determinants $\{|D_i\rangle\}$, each determinant corresponding to a given occupation by the 
$N_{\alpha}$ and $N_{\beta}$ electrons of $N=N_{\alpha}+N_{\beta}$ electrons
among a set of $M$ spin-orbitals $\{\phi_1,...,\phi_M\}$ (restricted case). When no symmetries are considered the maximum number of such determinants 
is ${M \choose N_{\alpha}}{M \choose N_{\beta}}$, a number that grows factorially with $M$ and $N$. 
The best representation of the exact wavefunction in the entire determinantal basis is the 
Full Configuration Interaction (FCI) wavefunction written as
\begin{equation}
|\Psi_0\rangle = \sum_i c_i |D_i\rangle
\end{equation}
where $c_i$ are the ground-state coefficients obtained by diagonalizing the Hamiltonian matrix, 
$H_{ij}=\langle D_i|H|D_j\rangle$, within the orthonormalized set, 
$\langle D_i|D_j\rangle= \delta_{ij}$, of determinants $|D_i\rangle$.\\
As well known, the main problem with FCI is the exponential increase of the wavefunction,
leading to unfeasible calculations except for small systems. However, an important feature of 
FCI expansion is that a vast majority of determinants 
have negligible coefficients due to their unphysical meaning and, in practice, only a tiny fraction of the FCI space is expected to be important.

To avoid handling the prohibitive size of the expansion one may try to select determinants by order of excitations with respect to the Hartree-Fock (HF) reference 
determinant, {\it e.g.} by taking all single and double excitations (CISD), triple and quadruple (CISDTQ) etc. but, here also, we are faced with 
the problem of handling a formidable number of determinants. For example, in the case of a CISD calculation the expansion size is 
about $(N_{\alpha}+N_{\beta})^2 n_{virt}^2$ where $n_{virt}$ is the number 
of virtual orbitals (unoccupied orbitals in the HF determinant), while for CISDTQ 
this size is of order $(N_{\alpha}+N_{\beta})^4 n_{virt}^4$. However, in both cases we are still managing unnecessarily a great number of determinants 
having a negligible weight in the expansion.

A natural idea to make such CI wavefunction much more compact (in practice, a most important property for DMC calculations) 
is to select among the FCI expansion only those determinants that contribute in a non-negligible way to the total energy. 
Such an idea and similar ones have been developed by several groups during the last decades 
(see, among others \cite{bender,cipsi1,buenker1,buenker2,buenker3,bruna,buenker-book,cipsi2,harrison}).
Here, we shall employ an approach close to that introduced by Huron {\it et al.}\cite{cipsi1} and Evangelisti {\it et al.}\cite{cipsi2}
Referred to as the CIPSI method (Configuration Interaction using a Perturbative Selection done Iteratively) it is based on a 
selection process constructed by using a perturbative estimate of the energy contribution of each determinant to a reference wave function
built iteratively. More details can be found in \cite{cipsi1,cipsi2}. Starting from this idea we have implemented a CIPSI-like algorithm to build compact 
trial wave functions to be used in FN-DMC calculations.\\

In its simplest form, the multi-determinant wavefunction is iteratively built as follows:\\

$\bullet$ Step 0: Start from a given determinant ({\it e.g.} the Hartree-Fock determinant) or set of determinants, thus defining 
an initial reference subspace: $S_0=\{|D_0\rangle,...\}$. Diagonalize $H$ within $S_0$ and get the ground-state energy $E_0^{(0)}$ and eigenvector:
\begin{equation}
|\Psi_0^{(0)}\rangle= \sum_{i \in S_0} c_i^{(0)} |D_i \rangle 
\end{equation}
Here and in what follows, the superscript on various quantities is used to indicate the iteration number.\\

Then, do iteratively ($n=0,...$):\\

{$\bullet$ Step 1}: Collect all {\it different} determinants $|D_{i_c}\rangle$ connected by $H$ to $|\Psi_0^{(n)}\rangle$, namely
\begin{equation}
\langle \Psi_0^{(n)}|H|D_{i_c}\rangle \ne 0
\end{equation}

{$\bullet$ Step 2}: Compute the second-order change to the total energy resulting from each connected determinant: 
\begin{equation}
\delta e(|D_{i_c}\rangle)=-\frac{{\langle \Psi_0^{(n)}|H|D_{i_c} \rangle}^2}{\langle D_{i_c}|H|D_{i_c}\rangle -E_0^{(n)}}
\label{e2pert}
\end{equation}

{$\bullet$ Step 3}: Add the determinant $|D_{i_c^*}\rangle$ associated with the largest $|\delta e|$ to the reference subspace: 
$$S_{n} \rightarrow S_{n+1}= S_{n} \cup \{|D_{i_c^*} \rangle\}$$\\

{$\bullet$ Step 4}: Diagonalize $H$ within $S_{n+1}$ to get: 
\begin{equation}
|\Psi_0^{(n+1)}\rangle= \sum_{i \in S_{n+1}} c_i^{(n+1)} |D_i\rangle \;\;\; {\rm with}\;\;\; E_0^{(n+1)}
\end{equation}

{$\bullet$} Go to step 1 or stop if the target size $N_{\rm dets}$ for the reference subspace has been reached.\\

Let us denote $|\Psi_0\rangle$ the wavefunction issued from the previous selection process and $E_0$ its variational energy. 
In all what follows, $|\Psi_0\rangle$ will be referred to as the CIPSI reference 
wavefunction and $E_0$ the variational CIPSI energy. 
Having constructed a zero-th order wavefunction, an improved estimate of the FCI energy can be obtained by adding to the variational energy 
the second-order correction, $E_{\rm PT2}$
\begin{equation}
E_{\rm PT2}=-\sum_{i \in M} \frac{  { {\langle \Psi_0|H|D_i} \rangle}^2}{\langle D_i|H|D_i\rangle - E_0}.
\label{ept2}
\end{equation}
where $M$ denotes the set of all determinants not belonging to the reference space and connected to the reference wavefunction $|\Psi_0\rangle$ 
by the Hamiltonian $H$ (single and double excitations). Finally, the total energy obtained is given by
\begin{equation}
E_0({\rm CIPSI})=E_0 + E_{\rm PT2}.
\label{ecipsitot}
\end{equation}
In this work, this latter energy will be referred to as the full CIPSI energy.\\

At this point a number of remarks are in order:\\

i.) Although the selection scheme is presented here for computing the ground-state eigenvector only, no special difficulties
arise when generalizing the scheme to a finite number of states (see, {\it e.g.}\cite{cipsi2})\\

ii.) The decomposition of the Hamiltonian $H$ underlying the perturbative second-order expression introduced in step 2 is given by 
$$H= H_0 + \langle D_{i_c}|H|D_{i_c}\rangle |D_{i_c}\rangle \langle D_{i_c}|$$
where $H_0$ is the restriction of $H$ to the reference subspace. This decomposition known as the Epstein-Nesbet 
partition\cite{en1,en2} is not unique, other possible choices are the M{\o}ller-Plesset partition\cite{mp} or the barycentric one,\cite{cipsi1} 
see discussion in \cite{cipsi2}.\\

iii.) Instead of calculating the energetic change perturbatively, expression (\ref{e2pert}), it can be preferable to employ
the non-perturbative expression resulting from the diagonalization of $H$ into the two-dimensional basis consisting of 
the vectors $|\Psi_0^{(n)}\rangle$ and $|D_{i_c}\rangle$. Simple algebra shows that the energetic change is given by
\begin{eqnarray}
\delta e(|D_{i_c}\rangle) &=&
      \frac{1}{2} \left[\langle D_{i_c}|H|D_{i_c}\rangle - E_0(N_{\rm dets})\right] \\
 & \times &
      \left[1-\sqrt{1 + \frac{4 {\langle \Psi_0^{(n)}|H|D_{i_c} \rangle}^2}{{[\langle D_{i_c}|H|D_{i_c}\rangle - E_0(N_{\rm dets})]}^2}}\right] \nonumber
\label{e2ex}
\end{eqnarray}
In the limit of small transition matrix elements, $\langle \Psi_0^{(n)}|H|D_{i_c}\rangle$, both expressions (\ref{e2pert}) and (\ref{e2ex}) coincide.
In what follows the non-perturbative formula will be used.\\

iv.) In step 3 a unique determinant is added at each iteration. Adding a few of them simultaneously is also possible, a feature 
particularly desirable when quasi-degenerate low-lying determinants are showing up. In the applications to follow this possibility has been systematically 
used by keeping at each iteration all determinants associated with an energetic change whose absolute value is greater than a given threshold.\\

v.) The implementation of this algorithm can be performed using limited amount of central memory. 
On the other hand, the CPU time required is essentially proportional to $N_{\rm dets} 
n_{\rm occ}^2 n_{\rm virt}^2$ where $n_{\rm occ}$ and 
$n_{\rm virt}$ are the number of occupied and virtual molecular 
orbitals, respectively.\\

\section{Application to the F$_2$ molecule}
\label{results_f2}
In this section calculations of the potential energy curve of the F$_2$ molecule both 
at the deterministic CIPSI and stochastic Fixed-Node DMC levels are presented. 
In subsection (\ref{cipsi_results}) we first present and discuss 
the results obtained with CIPSI. The dependence of the variational and full CIPSI energy 
curves on the number of selected determinants and on the basis set (Dunning's cc-pVDZ, 
cc-pVTZ, and aug-cc-pVTZ) is analyzed. Following the standard implementation of CIPSI, 
the number of determinants selected is kept constant along the potential energy curve; 
this approach is referred to 
as ``CIPSI at constant number of determinants''. In subsection (\ref{qmc_results}) 
FN-DMC results using CIPSI reference functions as trial wavefunctions are presented. 
To quantify the overall quality of the energy curves obtained either by CIPSI or FN-DMC 
we introduce in subsection (\ref{parallelerror}) the definition used here for the non-parallelism error 
measuring the degree of non-parallelism between the computed and exact curves.
Results for CIPSI and FN-DMC curves are given. To decrease the non-parallelism error without increasing 
the basis set size and, thus, the number of determinants that would 
make DMC calculations not feasible in practice, we propose a strategy based on the use of 
CIPSI reference functions with a variable number of determinants along the energy curve. For each geometry, 
the CIPSI selection process is stopped for a number of determinants leading to a given value 
of $E_{\rm PT2}$, Eq.(\ref{ept2}) common to all nuclear distances; this approach presented 
in subsection (\ref{ctept2_results}) is referred to as 
``CIPSI at constant $E_{\rm PT2}$''. FN-DMC results obtained with such trial wavefunctions 
are presented in subsection (\ref{fndmcept2}). Finally, a graphical summary of 
the potential energy curves obtained here at various levels of approximation is presented 
in subsection (\ref{graphi}).

\subsection{CIPSI at constant number of determinants}
\label{cipsi_results}
\begin{figure}[h]
\begin{center}
\includegraphics[width=0.9\columnwidth, angle=0, scale=1.0]{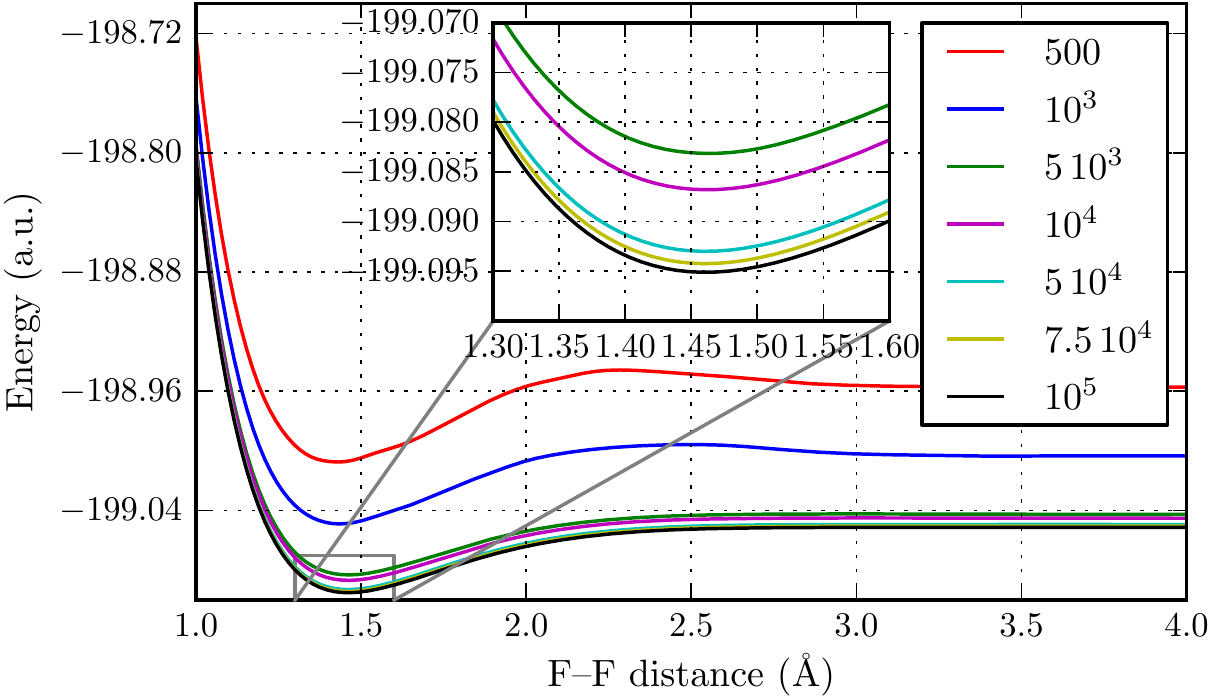}
\end{center}
\caption{cc-pVDZ basis set. Convergence of the variational CIPSI energy as a function of the number of 
selected determinants. Inset = blow up of the equilibrium region.}
\label{fig1}
\end{figure}

In figure \ref{fig1} variational CIPSI energy curves for increasing numbers of selected determinants 
are presented. The curves have been drawn from the energy values calculated at 27 interatomic distances. 
Interpolation between points is made using standard cubic splines.
The number of determinants is kept constant along the potential 
energy curve. The atomic basis set used is the Dunning cc-pVDZ (VDZ) basis set.\cite{dunning} 
For all basis sets considered in this work, CIPSI calculations are done with the molecular orbitals 
obtained from a minimal Complete-Active-Space Self-Consistent-Field (CASSCF) calculation 
(two electrons in two orbitals); no re-optimization of the 
molecular orbitals is performed. In addition, the core electrons are kept frozen. The curves of Fig. \ref{fig1} are obtained by stopping 
the CIPSI iterative process for a number of determinants $N_{\rm dets}$= 5 10$^2$, 10$^3$, 5 10$^3$, 
10$^4$, 5 10$^4$, 7.5 10$^4$, and 10$^5$. At the scale of the figure the 500- and 1000-determinant 
energy curves are not yet converged to the full CI. In both cases a fictitious dissociation barrier 
at intermediate internuclear distances is observed. This artefact associated with the lack 
of convergence of the multi-determinantal expansion disappears
for larger numbers of determinants. A convergence of the whole curve at the kcal/mol level
is reached for a number of determinants between 5 10$^4$ and  7.5 10$^4$.
Using the cc-pVDZ basis set (24 atomic orbitals) the size of the FCI space is about 
10$^{12}$ determinants (1$s$ orbitals frozen and no symmetry taken into account). 
The rapid convergence of variational CIPSI energy curves for such a large size 
illustrates the benefit of considering selected CI instead of more conventional CI schemes based 
on the use of full subspaces corresponding to 
multi-excitations of increasing order (all single-, all single- and double-, etc.) whose sizes 
become rapidly too large. Here, a CISD calculation (all single- and double- excitations) 
leads to a subspace size of about 2 10$^4$, while all SDTQ excitations (CISDTQ) 
generate about 10$^9$ determinants.

\begin{figure}[h]
\begin{center}
\includegraphics[width=0.9\columnwidth, angle=0, scale=1.]{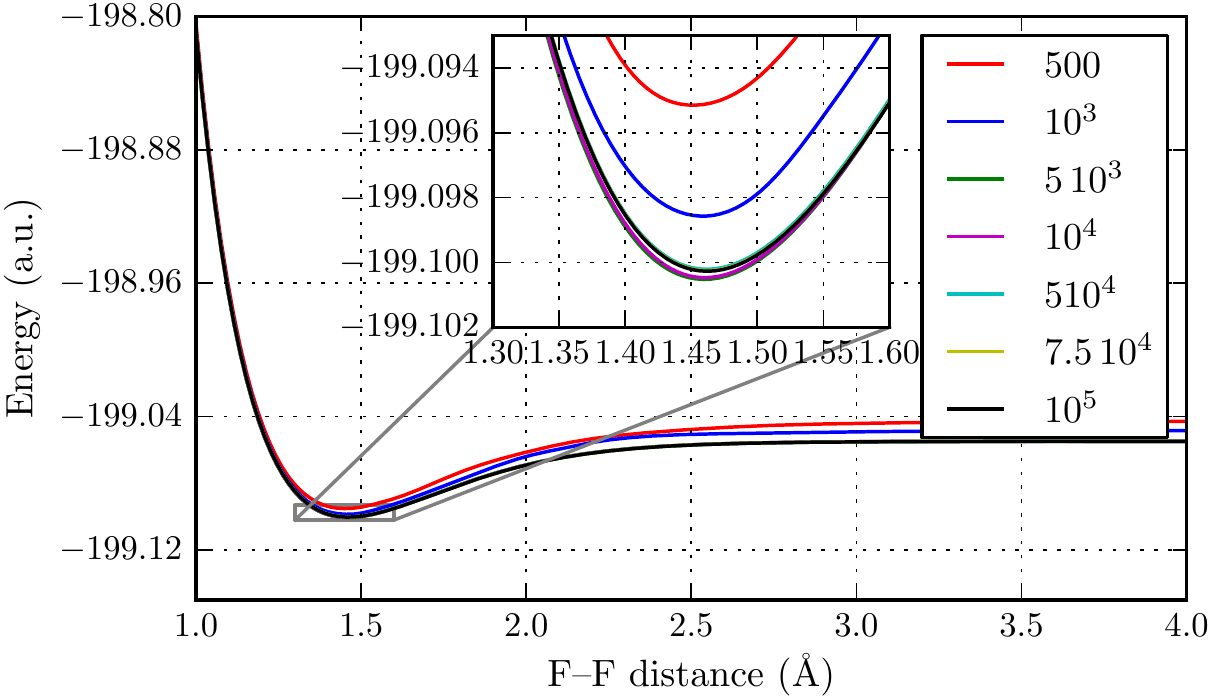}
\end{center}
\caption{cc-pVDZ basis set. Convergence of the full CIPSI energy as a function of the number of selected 
determinants. Inset = blow up of the equilibrium region.}
\label{fig2}
\end{figure}

Figure \ref{fig2} shows the full CIPSI energy curves obtained by adding to each variational energy 
the second-order perturbation energy correction, Eq.(\ref{ept2}). The improvement of the convergence 
in the number of determinants is striking. At about 5 10$^3$ determinants the convergence 
of the whole energy curve is reached with chemical accuracy ($\sim$ 1. kcal/mol).
At 10$^5$ determinants, the full CIPSI curve is expected to be an accurate estimate 
of the exact nonrelativistic full valence CI-VDZ potential energy curve. To quantify this latter aspect 
we report in Table \ref{tab1} our CIPSI energies together with the values of 
Bytautas {\it et al.} \cite{bytautas} calculated with the Correlation Energy Extrapolation by 
Intrinsic Scaling (CEEIS) approach. The CEEIS energies are believed to coincide 
with the exact FCI values with an accuracy of about 0.3 mEh. At the experimental 
equilibrium distance we also report the total energies obtained by 
Cleland {\it et al.} using {\it i}-FCIQMC\cite{booth2}.
The 13 internuclear distances of Table \ref{tab1} are those considered by Bytautas {\it et al.}. 
At the VDZ level the differences between CIPSI and CEEIS values are very small.
Around the equilibrium energy the maximum error between them is about 0.2 mEh and 
is slightly larger
at larger distances with a maximum of about 0.4 mEh.
At the equilibrium distance both CIPSI and CEEIS values almost coincide with the {\it i}-FCIQMC results.\\

\begin{figure}[h]
\begin{center}
\includegraphics[width=0.9\columnwidth, angle=0, scale=1.05]{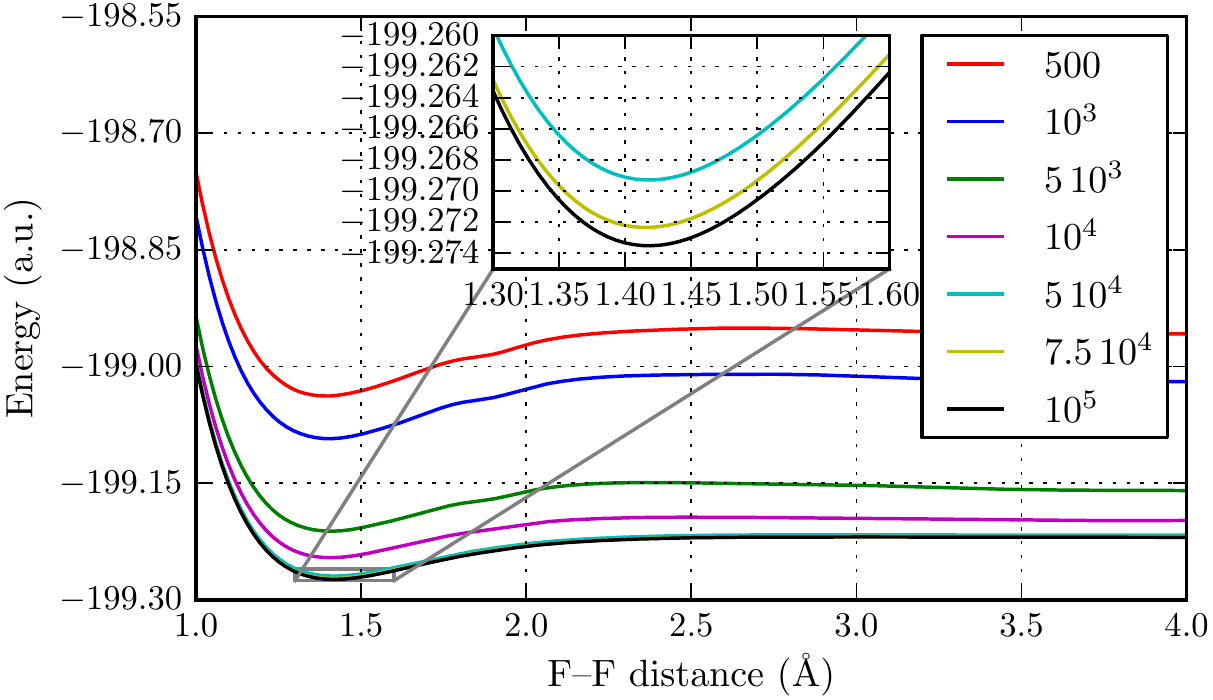}
\end{center}
\caption{cc-pVTZ basis set. Convergence of the variational CIPSI energy as 
a function of the number of determinants in the reference wave function. Inset = blow up 
of the equilibrium region.}
\label{fig3}
\end{figure}

\begin{figure}[h]
\begin{center}
\includegraphics[width=0.9\columnwidth, angle=0, scale=1.05]{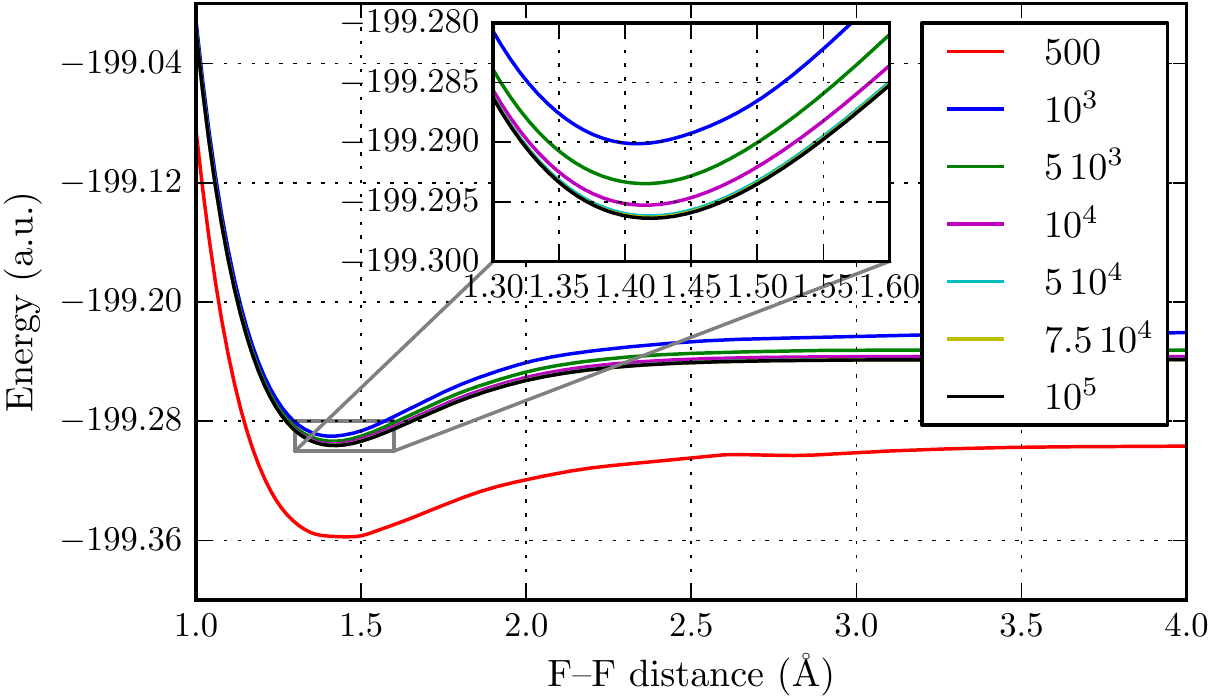}
\end{center}
\caption{cc-pVTZ basis set. Convergence of the full CIPSI energy as a function of 
the number of determinants. Inset = blow up of the equilibrium region.}
\label{fig4}
\end{figure}

\begin{figure}[h]
\begin{center}
\includegraphics[width=0.9\columnwidth, angle=0, scale=1.]{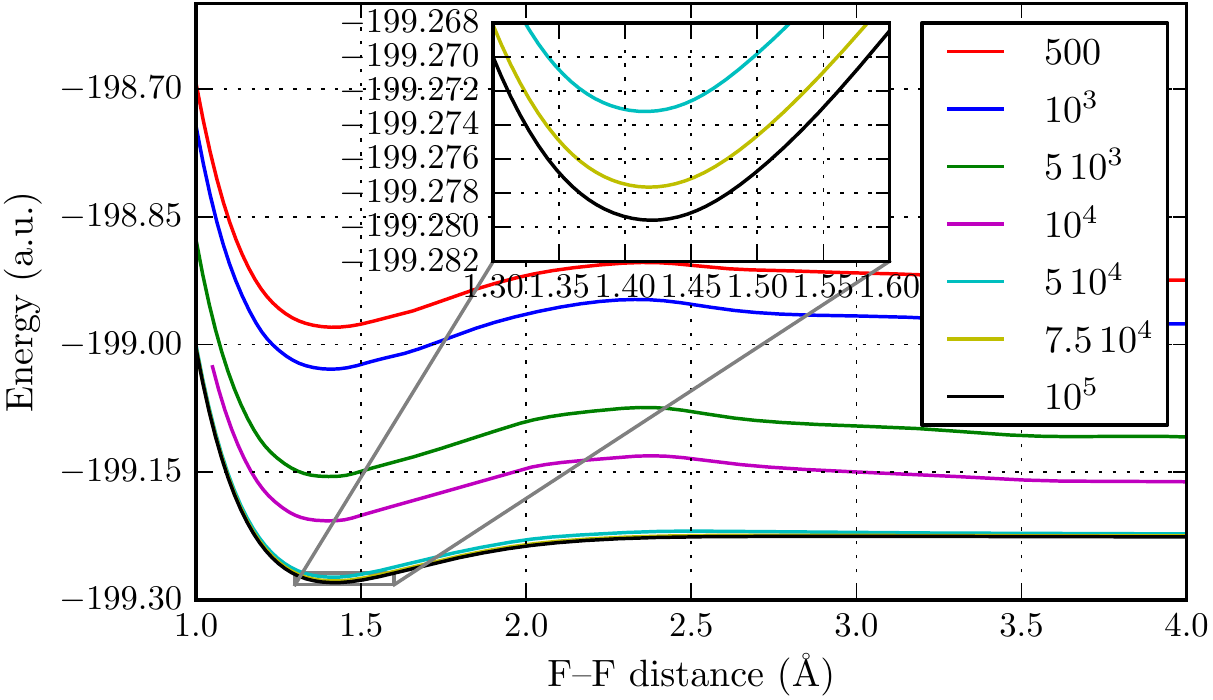}
\end{center}
\caption{aug-cc-pVTZ basis set. Convergence of the variational CIPSI energy as a function 
of the number of determinants. Inset = blow up of the equilibrium region.}
\label{fig5}
\end{figure}

\begin{figure}[h]
\begin{center}
\includegraphics[width=0.9\columnwidth, angle=0, scale=1.]{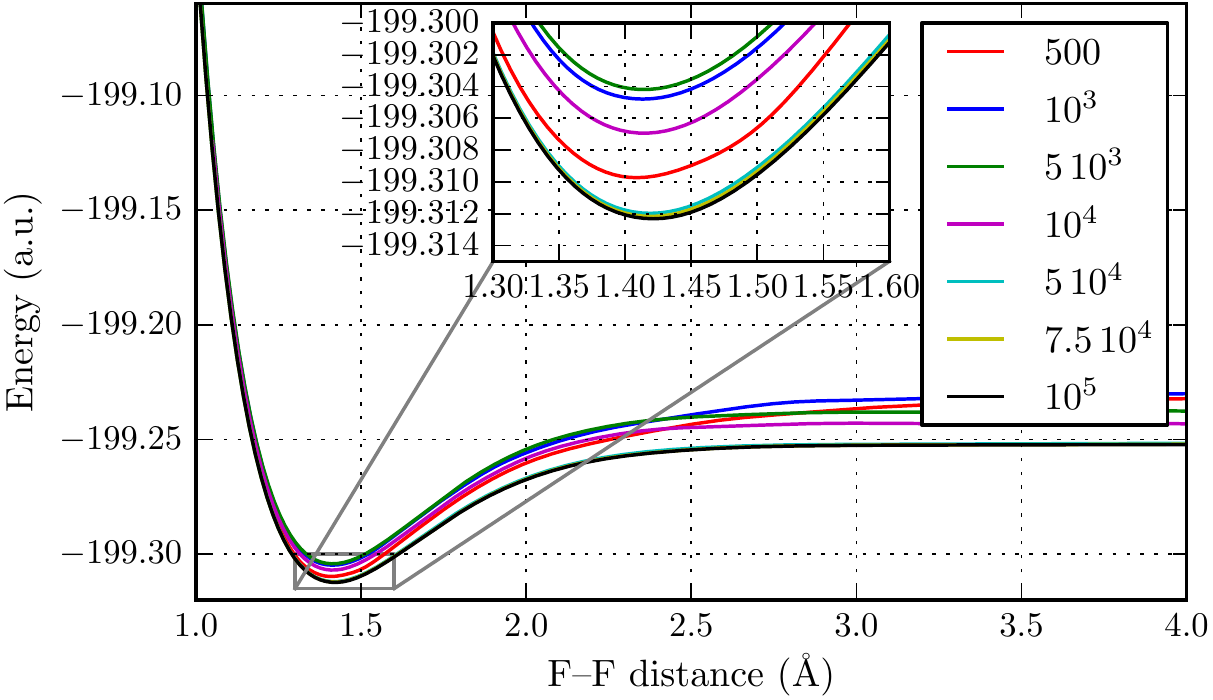}
\end{center}
\caption{aug-cc-pVTZ basis set. Convergence of the full CIPSI energy as a function of 
the number of determinants. Inset = blow up of the equilibrium region.}
\label{fig6}
\end{figure}

Figures \ref{fig3}, \ref{fig4}, \ref{fig5}, and \ref{fig6} present the results obtained with 
the greater cc-pVTZ (VTZ) and aug-cc-pVTZ (AVTZ) basis sets. The numbers of atomic orbitals are now 
60 and 92, respectively. The sizes of the Full CI space are much increased, about 10$^{20}$ and 
10$^{23}$ for the cc-pVTZ and aug-cc-pVTZ basis sets, respectively. 
As expected, the greater the Hilbert space is, the slower the convergence of the energy curves is. 
At the variational level, the convergence at the chemical level
with VTZ, Fig.\ref{fig3}, is attained for a number of determinants of about 10$^5$.
With AVTZ, Fig. \ref{fig5}, this level of convergence is not reached even with 10$^5$ determinants. 
As for the VDZ basis set, the convergence is greatly enhanced when the 
second-order perturbative correction is added up. The full CIPSI-VTZ energy curve 
is converged with a maximum error of about 1. kcal/mol for about 2 10$^4$
determinants, Fig.\ref{fig4}. As seen from Table \ref{tab1} the errors with respect to the accurate values 
of Ref. \cite{bytautas} are small. Around the equilibrium distance the error is of 
order 1. mEh and about two times larger in the long-distance regime. The convergence of the full CIPSI 
with the largest AVTZ basis set is still satisfactory and 
is obtained here at the kcal/mol level with a number of determinants greater than 4 10$^4$ determinants.
Once again, we emphasize that obtaining such good quality FCI curves
with a number of determinants representing only a tiny fraction of the whole 
Hilbert space is particularly remarkable (fractions of
about 10$^{-7}$, 10$^{-15}$, and 10$^{-18}$ for the VDZ, VTZ, and AVTZ basis sets, respectively).

\begin{table*}[t]
\begin{center}
\begin{tabular}{|l|c|c|c|c|c|c|c|}
\hline
\scriptsize
R in $\mathring{\rm A}$
& \scriptsize CIPSI-VDZ & \scriptsize CEEIS-VDZ$^b$ & \scriptsize {\it i}-FCIQMC-VDZ$^c$ & 
  \scriptsize CIPSI-VTZ & \scriptsize CEEIS-VTZ$^b$ & \scriptsize {\it i}-FCIQMC-VTZ$^c$ & 
\scriptsize CIPSI-AVTZ\\
\hline
1.14       & -199.007 16 & -199.007 18 &-             &-199.212 7  & -199.213 4 &-&-199.228 3\\
1.20       & -199.048 02 & -199.048 11 &-             &-199.252 2  & -199.253 0 &-&-199.267 9\\
1.30       & -199.084 94 & -199.085 10 &-             &-199.286 3  & -199.287 0 &-&-199.302 1\\
1.36       & -199.095 18 & -199.095 17 &-             &-199.294 2  & -199.295 0 &-&-199.310 2\\
1.41193$^a$& -199.099 28 & -199.099 20 &-199.099 41(9)&-199.296 5  & -199.297 2 &-199.297 7(1)&-199.312 3\\
1.50       & -199.099 77 & -199.099 81 &-             &-199.293 5  & -199.294 4 &-&-199.309 5\\
1.60       & -199.095 08 & -199.095 10 &-             &-199.285 2  & -199.286 1 &-&-199.301 2\\
1.80       & -199.080 90 & -199.080 90 &-             &-199.266 2  & -199.267 6 &-&-199.281 8\\
2.00       & -199.069 07 & -199.068 82 &-             &-199.252 7  & -199.254 3 &-&-199.267 4\\
2.20       & -199.061 84 & -199.061 65 &-             &-199.245 3  & -199.247 1 &-&-199.259 4\\
2.40       & -199.058 06 & -199.058 23 &-             &-199.241 7  & -199.243 6 &-&-199.255 6\\
2.80       & -199.055 52 & -199.055 77 &-             &-199.239 3  & -199.241 2 &-&-199.252 3\\
8.00       & -199.055 06 & -199.055 45 &-             &-199.238 4  & -199.240 8 &-&-199.250 0\\ 
\hline
\multicolumn{8}{|c|}{Atomic limit F+F}\\
\hline
\multicolumn{1}{|c}{}&\multicolumn{3}{c}{VDZ} &\multicolumn{3}{c}{VTZ} &\multicolumn{1}{c|}{AVTZ}\\
\multicolumn{1}{|l}{CIPSI, this work}&\multicolumn{3}{c}{-199.055 53}  &\multicolumn{3}{c}{-199.241 1}
&\multicolumn{1}{c|}{-199.255 9}\\
\multicolumn{1}{|l}{{\it i}-FCIQMC}&\multicolumn{3}{c}{-199.055 44(8)$^c$}
&\multicolumn{3}{c}{-199.241 0(2)$^c$}
&\multicolumn{1}{c|}{-}\\
\hline
\hline
\end{tabular}
\end{center}
\caption{Total nonrelativistic ground-state energies 
calculated using CIPSI (core electrons frozen). Basis sets= cc-pVDZ, 
cc-pVTZ, and aug-cc-pVTZ. For VDZ and VTZ results are compared with the values of 
Bytautas {\it et al.} \cite{bytautas} and those of Cleland {\it et al.} \cite{booth2} at the 
experimental distance. Energy in hartree.\\
$^a$ Expt. equilibrium distance\\
$^b$ Ref. \cite{bytautas}\\
$^c$ Ref. \cite{booth2}\\
}
\label{tab1}
\end{table*}

\begin{figure}[h]
\begin{center}
\includegraphics[width=0.9\columnwidth, angle=0, scale=1.05]{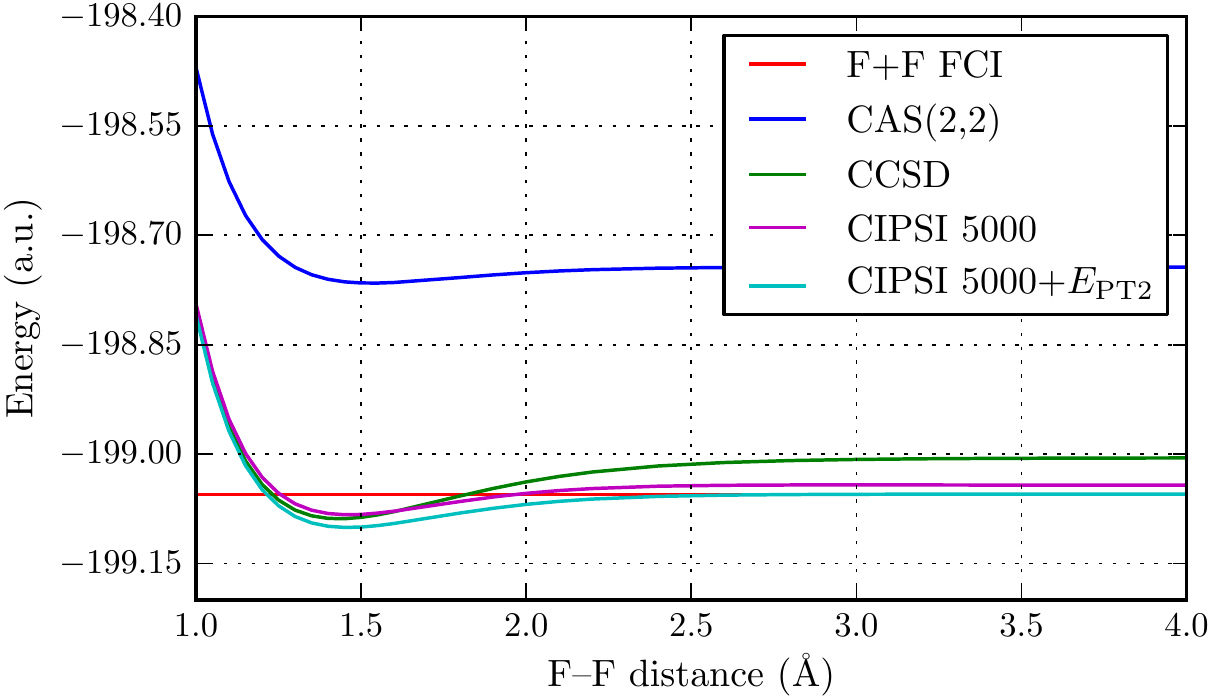}
\end{center}
\caption{cc-pVDZ basis set. Comparison of CIPSI results with CASSCF (upper part) and CCSD. The exact FCI (F+F) dissociation limit is given.}
\label{fig7}
\end{figure}

In figure \ref{fig7} a comparison of the CIPSI energy curves with those obtained 
by more standard approaches using the VDZ basis set is presented.
The CASSCF(2,2) potential energy curve is plotted in the upper part of the figure. 
Due to the absence of dynamical correlation contributions, CASSCF values are much too high in energy. 
However, as it should be the energy curve displays a correct 
dissociation behavior, the large-distance energy converging to the sum of HF energies 
of the two fluorine atoms. 
Note that the Hartree-Fock curve is not given here since at this level of approximation the 
F$_2$ molecule is not even
bound. The Coupled-Cluster curve (green line) using single and double excitations gives much 
more satisfactory results. In the equilibrium geometry region, CCSD energies are close to CIPSI results but still slightly higher. 
However, at large separations the CCSD curve dissociates with a large error of about 0.5 a.u with respect to the FCI-VDZ atomic energies.
This latter atomic limit is drawn on the figure as a horizontal line. 
As seen the variational CIPSI energy curve 
almost dissociates toward the exact value (error of about 0.013 a.u. at $R=4 \mathring{\rm A}$
a small but discernible quantity on the figure) while the full CIPSI energy curve 
is in full agreement with the FCI atomic limit (indistinguishable on the figure).

To get a more quantitative view of the dependence of the CIPSI potential energy curves 
on the number of determinants selected, we report the results obtained for the three basic spectroscopic 
quantities: The equilibrium distance, $R_{eq}$, the dissociation energy $D_0$, and the second 
derivative $k$ at $R_{eq}$ (curvature), a quantity directly related to the harmonic frequency. 
Data for the VDZ, VTZ, and AVTZ basis sets are given in Table \ref{tab2}
In this work, an accurate approximation of the exact non-relativistic, infinite nuclear masses, 
potential energy curve of the fluorine molecule has been built from data given 
in Refs.\cite{f2exact,f2exactII}. From $R=1.14 \mathring{\rm A}$ to $R=2.4 \mathring{\rm A}$ 
total energies are reconstructed from the non-relativistic contributions of Table IV 
in \cite{f2exact}, while for 
the long-range regime, from $R=2.8 \mathring{\rm A}$ to $8 \mathring{\rm A}$, 
the data used are taken from Table V of \cite{f2exactII}. The (F+F) dissociation limit is calculated 
using the atomic value of Davidson {\it et al.} in \cite{davidson_exact}. Note that the maximum 
error in the total energy curve resulting from these data 
is expected to be much smaller than the millihartree, a precision sufficient for our needs. 
To get the spectroscopic quantities, an accurate fit of the energies calculated at 26 
interatomic distances via a 10-parameter generalized Morse potential representation has been performed. 
In the case of the VDZ basis set we also report an accurate estimate of the VDZ dissociation energy 
obtained from the {\it i}-FCIQMC energy calculation of 
Cleland {\it et al.}\cite{booth2}. The value of $45.00(11)$ a.u reported in the table is a 
distance-corrected value that 
we have obtained by adding to the value of $43.87(11)$ a.u from Ref.\cite{booth2} calculated 
at the experimental equilibrium distance of $R= 1.4119 \mathring{\rm A}$,
the correction needed to shift to the minimum of the VDZ energy curve at $R=1.463 \mathring{\rm A}$.
As expected, CASSCF values for the three spectroscopic quantities are of low quality.
With the biggest AVTZ basis set, the CASSCF dissociation energy is less than one-half (28.4 mEh) 
of the exact value, the equilibrium distance is much too large, and the curvature too small. 
Using the VDZ basis set variational CIPSI results give an equilibrium distance and curvature 
essentially converged between $10^4$ and $5 10^4$ determinants. 
In contrast, the dissociation energy is still varying from $10^4$ ($D_0$= 41.9 mEh) 
to $10^5$ determinants ($D_0$= 43.97 mEh). At the full CIPSI level, 
the convergence of the dissociation energy is much better but still slightly decreasing. 
The value for $10^5$ determinants is $45.17$ mEh, to be compared 
with the corrected value of 45.00(11) mEh of Ref. \cite{booth2}. We also 
report the spectroscopic quantities obtained from the CEEIS data. The agreement with our own 
data is excellent, thus confirming that at the VDZ level (see, Table \ref{tab1}) 
the results obtained with CIPSI are of a quasi-FCI quality. 
Upon increasing the basis to the larger cc-pVTZ and aug-cc-pVTZ basis sets,
the spectroscopic quantities are significantly improved. The size of the Hilbert
spaces greatly increasing, the convergence is slowed down. At the VTZ level, the CIPSI
dissociation energy is still varying (decreasing) up to 5 $10^4$ determinants, 
the value obtained with the largest calculation being $D_0=57.6$ mEh.
Compared to the VDZ value of 44.1 mEh this value is much improved 
(exact value of 62.35 mEh). As it should be, the equilibrium distance is reduced and
the curvature increased as a result of the deepening of the well. 
The comparison with CEEIS data is also very good, except for the dissociation energy which 
we found about 1.mEh larger with CIPSI (57.6 and 56.7). This difference clearly results from 
the increase as a function the distance of the CIPSI error with respect to the quasi-FCI (CEEIS)
results: At  R=1.41193 the error is 0.7 mH and 2.4 mH at R=8. 
At the AVTZ level the CIPSI dissociation energy for $10^5$ determinants 
is found to be 60.0 mEh. However, because of the increase of the CIPSI error with distance
just discussed in the AVTZ case, this value must be taken with lot of caution and is very likely 
overestimated by one or two millihartrees. This is confirmed by the fact that using the 
CEEIS data of Bytautas {\it et al.},
\cite{f2exact} with the larger VQZ basis set the dissociation energy (expected to be 
larger than with AVTZ) is 59.8 mEh.

\begin{table*}[t]
\begin{center}
\begin{tabular}{|c|c|c|c|c|c|c|c|c|}
\hline
\hline
\multicolumn{9}{|c|}{VDZ basis set}\\
\hline
\multicolumn{1}{|c|}{ }& \multicolumn{1}{c|}{\footnotesize{CASSCF}}
&\multicolumn{5}{c}{\footnotesize{CIPSI: Variational$^a$/Full$^b$}}
&\multicolumn{1}{|c|}{\footnotesize{CEEIS$^c$}}
&\multicolumn{1}{|c|}{\footnotesize{$i$-FCIQMC$^d$}}\\
\hline
\footnotesize{$N_{\rm dets}$} & 2 & 5 $10^3$ & $10^4$ &5 $10^4$ & 7.5 $10^4$&$10^5$&  &\\
\hline
\footnotesize{$R_{eq}$}  & 1.531  &1.465/1.460&1.464/1.460&1.463/1.462& 1.462/1.463 
&1.463/1.463 & 1.460& -\\
\footnotesize{$D_0$ }    &  22.1  &40.9/45.7  &41.9/45.6&43.8/45.3&44.39/45.22& 43.97/45.17
&45.14 &45.00(11)\\
\footnotesize{$k$}       &  0.43  &0.73/0.78    &0.75/0.78  &0.77/0.77& 0.78/0.76 &0.76/0.76
&0.80&-\\
\hline
\hline
\multicolumn{9}{|c|}{VTZ basis set}\\
\hline
\multicolumn{1}{|c|}{ }& \multicolumn{1}{|c|}{\footnotesize{CASSCF}}
&\multicolumn{5}{|c|}{\footnotesize{CIPSI: Variational/Full}} 
& \multicolumn{2}{|c|}{\footnotesize{CEEIS}}\\
\hline
\multicolumn{1}{|c|}{\footnotesize{$N_{\rm dets}$}} & \multicolumn{1}{|c|}{2}  
& \multicolumn{1}{|c|}{5 $10^3$} &\multicolumn{1}{|c|}{$10^4$} &\multicolumn{1}{|c|}{5 $10^4$} 
& \multicolumn{1}{|c|}{7.5 $10^4$} & \multicolumn{1}{|c|}{$10^5$} & \multicolumn{2}{|c|}{}\\
\hline
\multicolumn{1}{|c|}{\footnotesize{$R_{eq}$}} 
& \multicolumn{1}{|c|}{1.469} & \multicolumn{1}{|c|}{1.410/1.412} &\multicolumn{1}{|c|}{1.409/1.415} 
&\multicolumn{1}{|c|}{1.418/1.417} &\multicolumn{1}{|c|}{1.418/1.419} 
&\multicolumn{1}{|c|}{1.418/1.417} &\multicolumn{2}{|c|}{1.416}\\
\multicolumn{1}{|c|}{\footnotesize{$D_0$}} & \multicolumn{1}{|c|}{ 26.5 } 
& \multicolumn{1}{|c|}{56.8/61.0}
&\multicolumn{1}{|c|}{49.8/58.6} &\multicolumn{1}{|c|}{52.9/57.6} 
&\multicolumn{1}{|c|}{49.8/58.6} &\multicolumn{1}{|c|}{54.14/57.6} 
& \multicolumn{2}{|c|}{56.7}\\
\multicolumn{1}{|c|}{\footnotesize{$k$}} & \multicolumn{1}{|c|}{ 0.63} 
& \multicolumn{1}{|c|}{1.067/1.132}
&\multicolumn{1}{|c|}{1.062/1.092} &\multicolumn{1}{|c|}{1.037/1.079} 
&\multicolumn{1}{|c|}{1.062/1.092} &\multicolumn{1}{|c|}{1.030/1.079} 
& \multicolumn{2}{|c|}{1.075}\\
\hline
\hline
\multicolumn{9}{|c|}{AVTZ basis set}\\
\hline
\multicolumn{1}{|c|}{ }& \multicolumn{1}{|c|}{\footnotesize{CASSCF}}
&\multicolumn{5}{|c|}{\footnotesize{CIPSI: Variational/Full}}& \multicolumn{2}{|c|}{\footnotesize{CEEIS}}\\
\hline
\multicolumn{1}{|c|}{\footnotesize{$N_{\rm dets}$}} & \multicolumn{1}{|c|}{2} 
& \multicolumn{1}{|c|}{5 $10^3$} & \multicolumn{1}{|c|}{$10^4$}
&\multicolumn{1}{|c|}{5 $10^4$} &\multicolumn{1}{|c|}{7.5 $10^4$}
&\multicolumn{1}{|c|}{$10^5$} & \multicolumn{2}{|c|}{}\\
\hline
\multicolumn{1}{|c|}{\footnotesize{$R_{eq}$}} 
& \multicolumn{1}{|c|}{1.463} & \multicolumn{1}{|c|}{1.405/1.413}
&\multicolumn{1}{|c|}{1.392/1.414} &\multicolumn{1}{|c|}{1.415/1.418} 
&\multicolumn{1}{|c|}{1.418/1.419} &\multicolumn{1}{|c|}{1.418/1.419}
& \multicolumn{2}{|c|}{-}\\
\multicolumn{1}{|c|}{\footnotesize{$D_0$}} & \multicolumn{1}{|c|}{ 28.4} & \multicolumn{1}{|c|}{59.6/66.4}
&\multicolumn{1}{|c|}{57.5/63.8} &\multicolumn{1}{|c|}{52.1/60.2}
&\multicolumn{1}{|c|}{54./60.0} 
&\multicolumn{1}{|c|}{54./60.0} 
& \multicolumn{2}{|c|}{-}\\
\multicolumn{1}{|c|}{\footnotesize{$k$}} & \multicolumn{1}{|c|}{ 0.66} & \multicolumn{1}{|c|}{1.158/1.151}
&\multicolumn{1}{|c|}{1.108/1.117}&\multicolumn{1}{|c|}{1.013/1.078} 
&\multicolumn{1}{|c|}{1.03/1.073} 
&\multicolumn{1}{|c|}{1.03/1.073} 
& \multicolumn{2}{|c|}{-}\\
\hline
\hline
\multicolumn{9}{|c|}{Infinite basis set (Exact)}\\
\hline
\multicolumn{1}{|c|}{\footnotesize{$R_{eq}$}} 
& \multicolumn{8}{|c|}{1.412}\\
\multicolumn{1}{|c|}{\footnotesize{$D_0$}} 
& \multicolumn{8}{|c|}{62.35$^e$}\\
\multicolumn{1}{|c|}{\footnotesize{$k$}} 
& \multicolumn{8}{|c|}{1.121}\\
\hline
\hline
\end{tabular}
\end{center}
\caption{Spectroscopic constants from variational and full CIPSI 
as a function of $N_{\rm dets}$ and of the basis 
set employed.
$R_{eq}$ in $\mathring{\rm A}$,  $D_0$ in millihartrees,
and $k$ in hartree/${\mathring{\rm A}}^2$.
Spectroscopic constants ($X=R_{eq}$, $D_0$, and $k$) presented as $X_1/X_2$ 
where $X_1$ and $X_2$ are the values obtained from the 
variational and full CIPSI energy curves, respectively.
For comparison, the CASSCF, exact non-relativistic, and the corrected near-FCI value of Cleland 
{\it et al.} (see, text) are also given.\\
$^a$ Spectroscopic data obtained from the variational CIPSI energy curve.\\
$^b$ Spectroscopic data obtained from the full CIPSI energy curve, Eq.(\ref{ecipsitot})\\
$^c$ \cite{bytautas}\\
$^d$ \cite{booth2}\\
$^e$ Value taken from Table IV of Bytautas {\it et al.} \cite{f2exact} and corresponding to 
the estimated nonrelativistic full valence CI dissociation energy (no core correlation). 
}
\label{tab2}
\end{table*}

At this point it would be desirable to increase further the basis set by considering higher 
cardinal numbers (QZ, 5Z, etc.). However, our purpose being to avoid to handle exponentially increasing 
Hilbert spaces, we shall now turn our attention to FN-DMC methods.

\subsection{FN-DMC with CIPSI at constant number of determinants}
\label{qmc_results}
We present all-electron FN-DMC calculations of the potential energy curve using CIPSI 
reference wavefunctions as trial wavefunctions. 
All FN-DMC simulations presented in this section have been performed by running first a 
CIPSI calculation at constant number of determinants and, then,
using as trial wavefunction the CIPSI reference wave function {\it as it is}. 
We emphasize that no preliminary multi-parameter stochastic optimization of 
the trial wavefunction has been performed, the CASSCF molecular orbitals are 
kept unchanged for all distances and calculations, the determinantal coefficients 
are those issued from CIPSI, and no Jastrow prefactor has been employed (we just impose 
the electron-nucleus cusp conditions at very short distances).
In this way all calculations may be performed 
in a fully automatic way: i) Choose a target number of determinants 
and of Monte Carlo steps, ii) run CIPSI, and then iii) run FN-DMC.

\begin{figure}[h]
\begin{center}
\includegraphics[width=0.9\columnwidth, angle=0, scale=1.05]{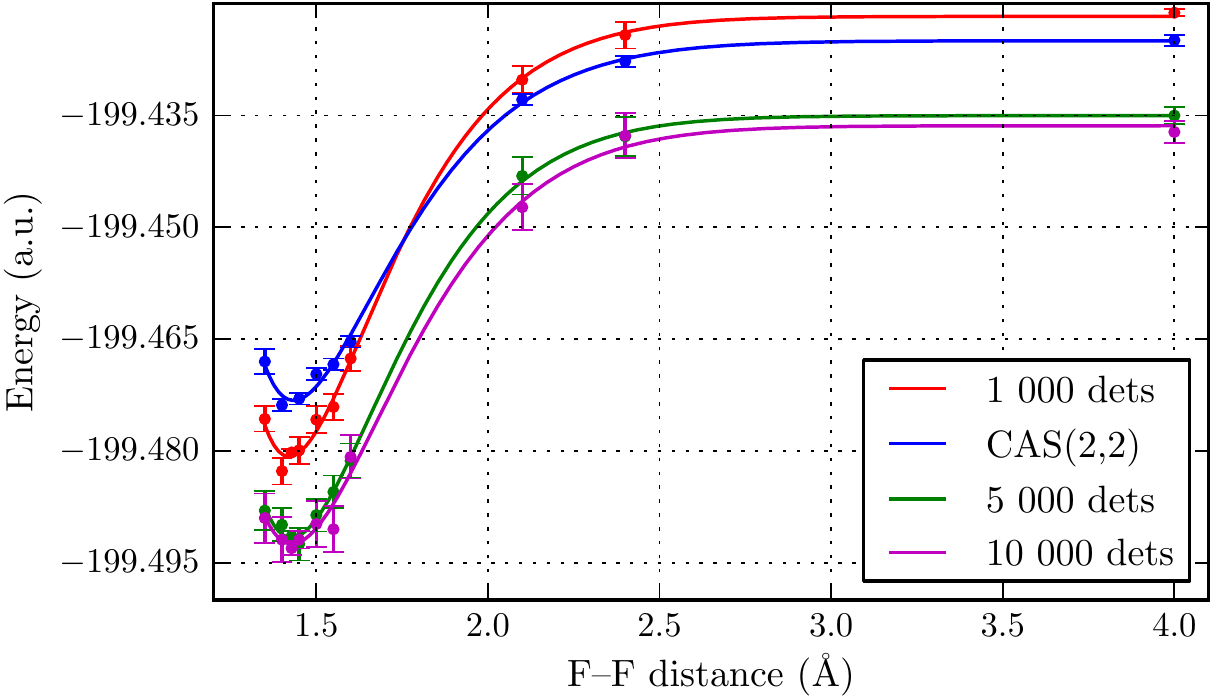}
\end{center}
\caption{FN-DMC with CIPSI-VDZ nodes. Convergence of the FN-DMC energy curve as a function 
of the number of determinants selected in the trial wavefunction}
\label{fig8}
\end{figure}

The convergence of the FN-DMC energy curve as a function of the number of determinants 
for the cc-pVDZ basis set is presented in figure \ref{fig8}. For the sake of comparison, 
the DMC curve obtained from CAS(2,2) nodes is also reported.
Total energies have been calculated for interatomic distances of 1.35, 1.40, 1.428, 
1.45, 1.50, 1.55, 1.60, 2.10, 2.40, and 4.00 $\mathring{\rm A}$. 
The statistical error on FN-DMC/CIPSI energies using 1 000 and 5 000 determinants 
is typically 0.002 a.u. For 10 000 determinants a slighter larger value of about 0.003 a.u
is obtained. To better estimate the dissociation 
energy (see, below) the energy values at the distances of 1.428 and 4. have 
been computed using more statistics (Monte Carlo runs ten times longer).
Quite remarkably, FN-DMC/CIPSI energies are found to systematically decrease for all 
interatomic distances as the number of selected determinants is increased 
Said differently, the nodes of the CIPSI wavefunctions are 
systematically improved upon iterations (reduction of the fixed-node error). 
A similar property has been observed for the oxygen atom\cite{canadian} and, also, 
for bigger atoms and molecules.\cite{preprint1,preprint2} To understand the origin 
of this remarkable mathematical property is not simple. However, a heuristic 
argument can be given as follows. When the total energy is lowered 
(the criterion used during the CIPSI selection process) 
the wavefunction is dominantly improved in the neighborhood of its maxima 
(regions contributing the most to the energy). In particular, the localization of such 
maxima is expected to be improved. Now, the positions of the maxima and zeroes of 
a wavefunction being intimately correlated (as any solution of a wave-like equation),
we can also expect an improved localization of the nodes.
As seen on the figure the convergence of the FN-DMC energy curve 
is approximately reached for 5 $10^3$ determinants, a result coherent 
with the convergence of the variational CIPSI energy curve 
at roughly the same number of determinants, Fig.\ref{fig1}.
Note that handling a few thousands of determinants in FN-DMC 
(the trial wavefunction and its derivatives are to be computed at each 
Monte Carlo step) is still feasible in practice.

In Table \ref{tab3} the FN-DMC spectroscopic quantities ($R_{eq}$, $D_0$, $k$) 
obtained with CAS(2,2) and CIPSI/VDZ nodes for increasing numbers of determinants
are presented. The values and error bars have been obtained 
by fitting a set of 20 energy curves using a 10-parameter generalized Morse potential 
representation. Each of these curves is obtained from different realizations of 
the statistical noise. For $D_0$ we also give the value obtained 
by directly computing the difference $\Delta E$ between total FN-DMC energies at the 
equilibrium geometry and at the large-distance value of R=4 $\mathring{\rm A}$. 
Spectroscopic quantities are essentially converged within error bars for 
5 $10^3$ determinants. However, due to the magnitude of statistical fluctuations, 
to extract accurate values of the equilibrium distances and curvatures is impossible. 
Using CIPSI nodes, typical values are $R_{eq}$= 1.43(3) and $k=$1.0(4). 
Error bars are clearly too large to allow any detailed analysis.
In contrast, the situation is more satisfactory for the dissociation energy.
Coherent values with small enough statistical errors are obtained 
either from the statistical fit of the energy curves or the direct 
calculation of the energy gap. Using 10$^4$ determinants the FN-DMC dissociation 
energy obtained is 55.2(12) mEh, a clear improvement with respect to the 
value of 41.9 mEh corresponding to the trial wavefunction (see, Table \ref{tab2})
Finally, remark that results obtained with FN-DMC/CIPSI-VDZ are 
of a comparable quality to those obtained at the variational CIPSI/VDZ 
level (see, Table \ref{tab2})

To improve the quality of the results beyond FN-DMC/CIPSI-VDZ, 
a natural solution is to increase 
the size of the basis set. However, such a strategy is doomed to failure since 
the number of determinants necessary to build converged nodes is expected 
to increase too rapidly. For example, 
in the case of the VTZ basis set for which the variational energy 
was found to converge around 5 $10^4$ determinants, a similar number 
of determinants should be 
expected to get accurate enough nodes. 
From a computational point of view, this situation is clearly not 
favorable to FN-DMC. 
In what follows, we want to avoid to follow this path and, instead, propose 
an alternative strategy 
based on the improvement of the global shape of the energy curve 
instead of on the search for increasingly precision for total energies. 

\begin{table*}[t]
\begin{center}
\begin{tabular}{|l||c|c|c|l|l|l|}
\hline
 &Exact &CAS-nodes& 10$^3$-det nodes& 5 10$^3$-det nodes
 & 10$^4$-det nodes\\
\hline
\hline
$R_{eq}$&1.412&1.434(8)$\;\;\;\;$&1.419(15)$\;\;\;$&1.424(26)$\;\;\;$&1.428(20)$\;\;\;$\\
\hline
$D_0$&62.0$^a$ &48.3(7)/$\Delta E$=48.3(10)&59.6(13)/$\Delta E$=59.0(6)
&56.9(16)/$\Delta E$=56.5(13)&55.2(12)/$\Delta E$=55.9(18)\\
\hline
$k$     &1.121 &1.0(2)$\;\;\;\;$&1.6(6) $\;\;\;\;$& 1.1(5) $\;\;\;\;\;\;$ 
& 1.0(4) $\;\;\;\;\;\;$ \\
\hline
\hline
\end{tabular}
\end{center}
\caption{FN-DMC spectroscopic quantities computed with CAS(2,2)
and CIPSI-VDZ nodes (number of determinants = 10$^3$, 5 $10^3$, and 10$^4$).
Values of ($R_{eq}$, $D_0$, $k$) and error bars obtained 
from a statistical distribution of 20 different energy curves fitted with a 
generalized Morse potential.
In the case of $D_0$, the dissociation energy directly obtained from the energy gap is 
also given. Equilibrium 
distances in $\mathring{\rm A}$, dissociation energies in millihartrees,
and energy curvatures in hartree/${\mathring{\rm A}}^2$\\
$^a$ Value taken from Table IV of Bytautas {\it et al.} \cite{f2exact} and corresponding to
the estimated non-relativistic full CI dissociation energy including the contribution of
the core correlation (present in FN-DMC).
}
\label{tab3}
\end{table*}

\subsection{Quantifying the non-parallelism error}
\label{parallelerror}
Along a potential energy curve -more generally a potential energy surface (PES)- the physical/chemical 
nature of the wavefunction is known to change dramatically. Thus, one of the critical issue 
of any electronic structure approach is its ability of treating with a similar precision 
the various regimes of the wavefunction (so-called ``balanced'' description of the PES). 
For chemical/physical purposes to get accurate estimates of absolute 
total energies is known not to be of great interest; instead, we 
need to accurately calculate the {\it variation} of the total energy (energy gradient) 
along a reaction path. To quantify such an aspect it is usual to introduce a quantity (the so-called 
non-parallelism error) measuring the 
degree of non-parallelism of the computed curve with the exact curve. 
This quantity being heuristic in nature, 
several definitions are possible. Here, the non-parallelism error, $\Delta$, is defined as follows. 
Denoting $\{E^{i}\}_{i=1,N}$ a set of $N$ approximate total energies 
computed with a given method for $N$ geometries, $\Delta$ is defined as
\begin{equation}
\Delta= \sqrt{ \frac{1}{N} \sum_{i=1}^N [E^i- (E_{ex}^{i}+\bar{d})]^2  },
\end{equation}
where $E_{ex}^{i}$ are the exact energies and $\bar{d}$ the average 
distance between the exact and the approximate curve
\begin{equation}
\bar{d}= \frac{1}{N} \sum_{i=1}^N |E^i-E_{ex}^{i}|
\end{equation}
Roughly speaking, $\Delta$ can be viewed as a measure of the variations of the computed curve 
with respect of the exact one, once the exact 
curve has been shifted upward by the average distance between the two curves. 
As it should be, $\Delta$ vanishes when the 
two curves are exactly parallel and increases as a function of the difference between 
the overall global shape of the two PES.

Table \ref{tab4} reports the values obtained for $\Delta$ at the various levels 
of approximation discussed up to now as a function 
of the number of determinants. To calibrate our data, the CASSCF-VDZ value, $\Delta=0.02371$, is 
given. At the CIPSI-VDZ level, the values corresponding to the variational and full CIPSI calculations
are found to converge to the value $\Delta \sim 0.011$. As discussed above, a quasi convergence 
to the FCI curve being obtained, this value 
should be considered as the non-parallelism error of the FCI-VDZ curve. 
As expected, using the larger VTZ basis set decreases 
the error. The value found for $\Delta$ is about 0.0035, a definite improvement 
with respect to the VDZ basis set. 
Calculating the energy curve with FN-DMC using CIPSI-VDZ reference function is also expected 
to decrease the non-parallelism error. 
The value obtained at this level is, $\Delta \sim 0.0028$, a value slightly smaller 
than the one found with the purely deterministic CIPSI-VTZ approach.

\begin{table*}[t]
\begin{center}
\begin{tabular}{|c|c|c|c|c|c|c|c|c|}
\hline
Number of determinants& 2 &10$^2$ & 5 10$^2$ & 10$^3$ & 5 10$^3$ & 10$^4$ & 5 10$^4$ & 10$^5$ \\
\hline
CASSCF-DZ  &0.0265& & & && && \\
CASSCF-TZ  &0.0204& & & && && \\
Variational CIPSI-VDZ&&0.0050& 0.0065& 0.0080& 0.0125& 0.0122& 0.0116& -\\
Full CIPSI-VDZ&&0.0071& 0.0087& 0.0093& 0.0110& 0.0110& 0.0112& -\\
Variational CIPSI-VTZ&&0.0054& 0.0072& 0.0092& 0.0108& 0.0101& 0.0056& 0.0047\\
Full CIPSI-VTZ&&0.0095& 0.0065& 0.0060& 0.0039& 0.0038& 0.0038& 0.0038\\
FN-DMC/CIPSI-VDZ&&0.0021& 0.0029& 0.0020& 0.0027& 0.0028& - & -\\
\hline
\end{tabular}
\end{center}
\caption{Non-parallelism error of the variational CASSCF curves, the variational and full 
CIPSI curves, and the FN-DMC curve with CIPSI-VDZ nodes.}
\label{tab4}
\end{table*}

To reduce further the non-parallelism error without increasing the basis set, an approach aiming 
at describing in a more coherent way the different regions of the potential energy curve 
is proposed in the following section.

\subsection{CIPSI at constant $E_{\rm PT2}$}
\label{ctept2_results}
Stopping the iterative CIPSI process at a given number of determinants identical for all geometries 
does not insure a coherent description of the energy curve. To construct a wavefunction 
of comparable quality in each region (short interatomic distances, 
equilibrium region, intermediate regime, and near-dissociation limit), 
it is natural to consider expansions involving a variable number of determinants 
as a function of the interatomic distance. This point is clearly illustrated 
on Fig.\ref{fig1} where the intermediate region between $R=2 \mathring{\rm A}$ 
and $2.5 \mathring{\rm A}$ is poorly described with a CIPSI-VDZ wavefunction having a
small number of determinants.
For 5 $10^2$ and $10^3$ determinants, a spurious local maximum is observed, this artefact 
disappearing for larger numbers of determinants. In contrast, in the equilibrium regime where 
the wavefunction is known to have a much less marked multi-configurational character, 
no qualitative change for the energy curve is observed when passing from a (very) small 
to a large number of determinants. Fig.\ref{fig2} shows that after adding the second-order correction 
to the variational CIPSI energy, all curves become much better-behaved and in particular 
no spurious maxima are observed. As already pointed out, this result shows that 
$E_{\rm PT2}$ represents most of the remaining difference between the variational CIPSI 
and FCI energies. It thus motivates us to consider the magnitude of $E_{\rm PT2}$ as an indicator 
for evaluating the difference between the CIPSI multi-determinantal 
wavefunction and the FCI solution. The new strategy is then as follows: At a 
given geometry the CIPSI iterative process is stopped when a target value for the second-order 
contribution is reached and not when a fixed number of determinants is obtained. 
The target value is chosen identical along the PES 
and the FCI limit is recovered when this value is decreased down to zero. 
When $E_{\rm PT2}$ represents a good estimate of the difference between the variational CIPSI and FCI 
energies, the variational CIPSI curve obtained with the variable number of determinants 
is expected to be almost parallel to the unknown FCI one. 
Of course, when higher-order corrections E$_{PTN}$ (N $>$ 2) 
are present, some deviation from the FCI curve is expected. 
Remark that some work with a similar idea has been done previously 
by Persico {\it et al.}\cite{persico}. The difference with the present work is that 
the norm of some approximate first-order perturbed wavefunction 
was considered instead. We believe that using directly an energetic criterion
is more natural in this context.

\begin{figure}[h]
\begin{center}
\includegraphics[width=0.9\columnwidth]{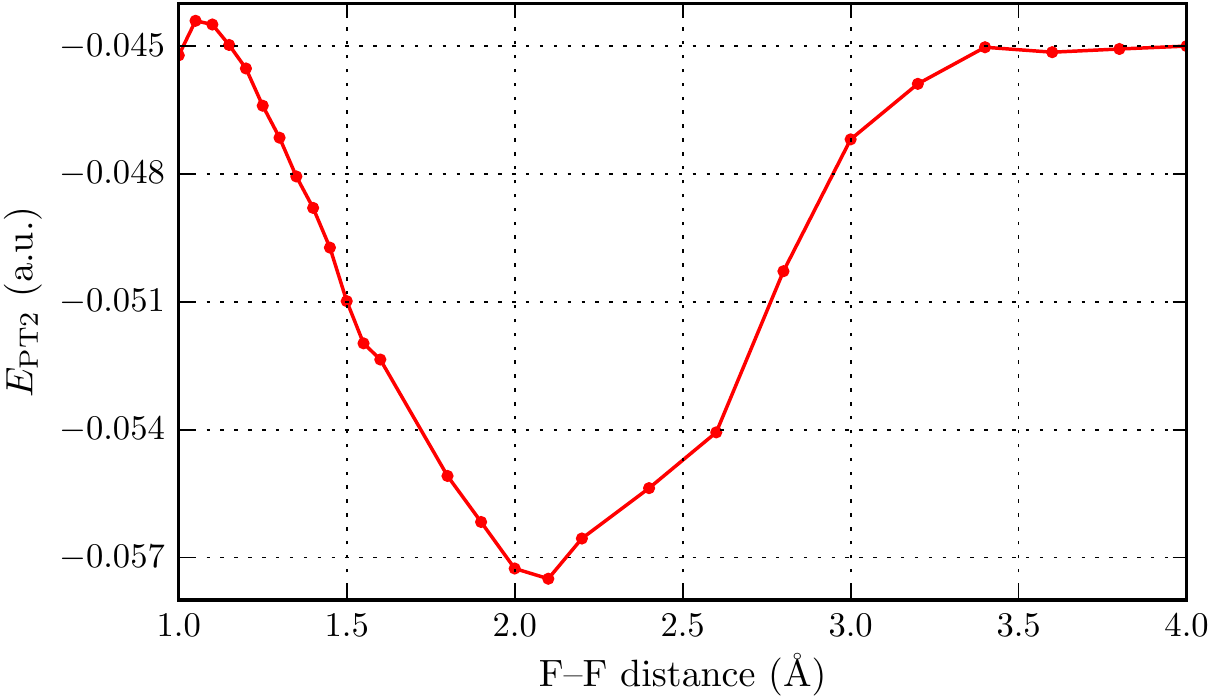}
\end{center}
\caption{$E_{\rm PT2}$ as a function of the interatomic distance 
for CIPSI-VDZ wavefunctions with 10$^3$ determinants. Basis set= cc-pVDZ}
\label{fig9}
\end{figure}

To illustrate how using a constant number of determinants along the potential energy curve 
can induce distortion in the quality 
of the reference function, we present in Fig. \ref{fig9} the second-order perturbative 
correction calculated 
from the CIPSI-VDZ reference wavefunction including 10$^3$ determinants 
as a function of the interatomic distance. 
In the intermediate region where the bond is about to be broken (around 2.2 $\mathring{\rm A}$) 
the perturbative correction is bigger that for any other geometries and more determinants 
are needed to produce a wavefunction of similar quality.

\begin{figure}[h]
\begin{center}
\includegraphics[width=0.9\columnwidth]{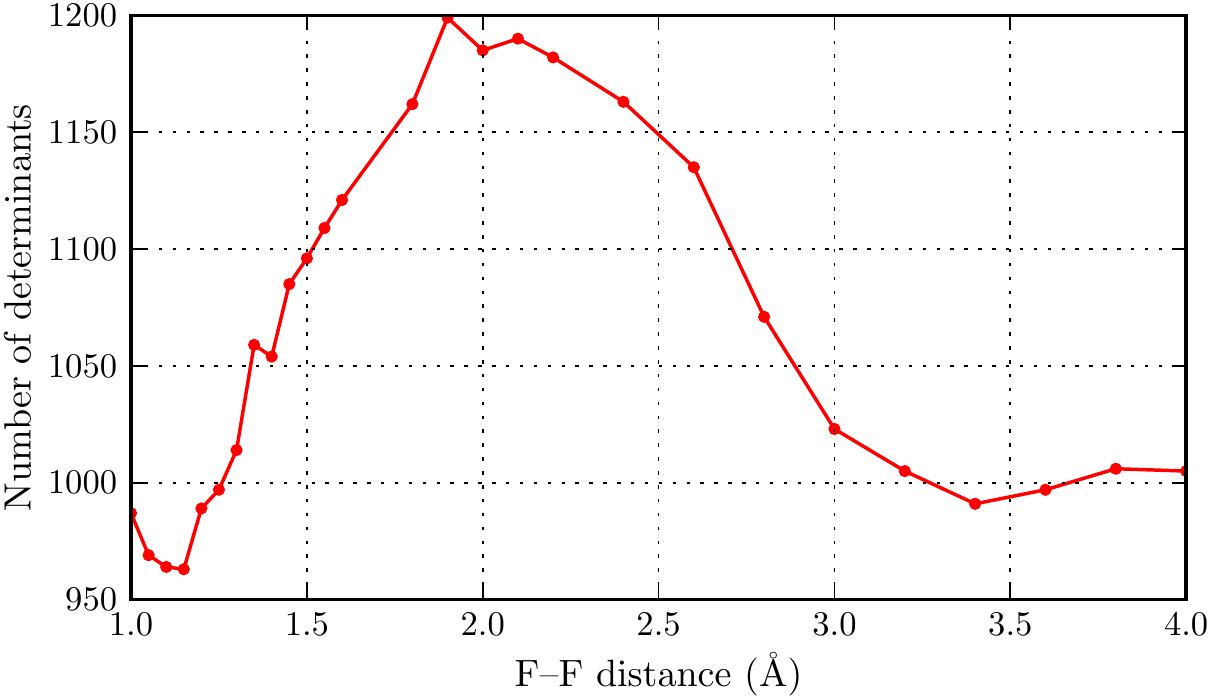}
\end{center}
\caption{Number of determinants required to impose a value of $E_{\rm PT2}$ of -0.05 hartree 
as a function of the interatomic distance. Basis set=cc-pVDZ} 
\label{fig10}
\end{figure}

In figure \ref{fig10} the number of determinants 
needed along the energy curve to impose a constant value of $E_{\rm PT2}$ of -0.05 hartree
is plotted. It is approximately the value obtained when considering a fixed number of 10$^3$ 
determinants, see Fig.\ref{fig9}. 
As it should be a non-constant value of the number of determinants is observed 
with a maximum in the intermediate region.

\begin{figure}[h]
\begin{center}
\includegraphics[width=0.9\columnwidth]{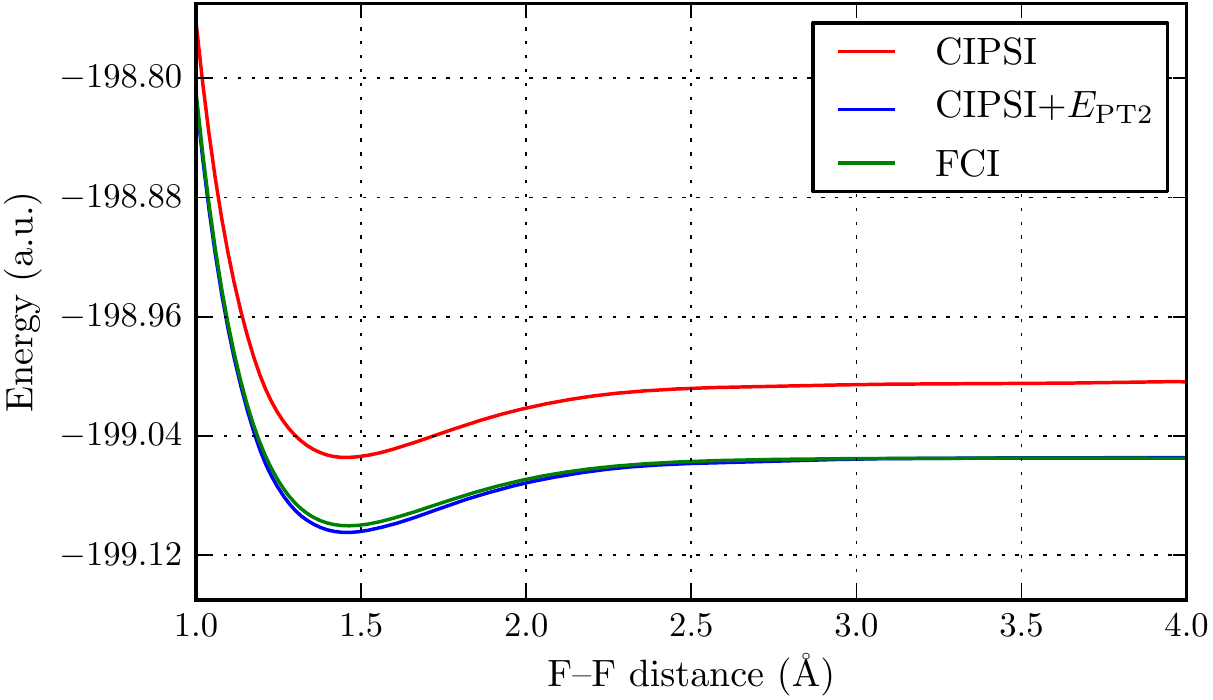}
\end{center}
\caption{Potential energy curves obtained by imposing a constant value $E_{\rm PT2}$ of -0.05 hartree.}
\label{fig11}
\end{figure}

The potential energy curves obtained using this scheme are presented in Fig.\ref{fig11} 
for a value of $E_{\rm PT2}$ of -0.05 hartree. The variational CIPSI energy curve obtained 
with a variable number of determinants (upper curve), 
the full CIPSI curve obtained by adding the constant perturbative contribution 
of -0.05, and the FCI one accurately approximated 
by the 10$^5$-determinant curve presented in Fig.\ref{fig2} are plotted. 
Remarkably, the FCI and full CIPSI curves are almost identical. It means 
that higher-order perturbative contributions 
beyond the second-order are small. Their largest contributions 
lie in the equilibrium region but do not exceed 0.005 a.u. 
Imposing a constant $E_{\rm PT2}$ thus leads to a full CIPSI curve 
close to the FCI one and then to a variational CIPSI energy curve 
almost parallel to the FCI one. 
Note also that the spurious maximum observed when using 
a constant small number of determinants is no longer present. 

\begin{figure}[h]
\begin{center}
\includegraphics[width=0.9\columnwidth]{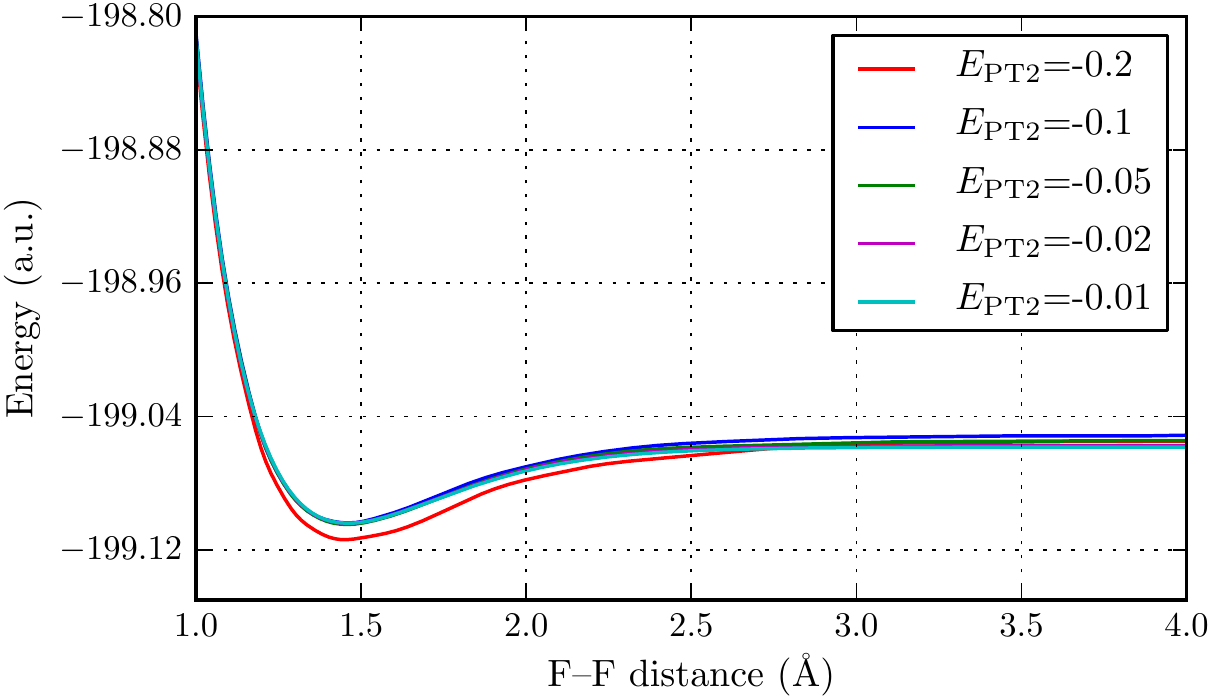}
\end{center}
\caption{Convergence of the full CIPSI energy curve for various values of constant $E_{\rm PT2}$.}
\label{fig12}
\end{figure}

Figure \ref{fig12} presents the full CIPSI curves obtained with 
$E_{\rm PT2}$=-0.2,-0.1,-0.05,-0.02, and -0.01 (atomic units). 
As seen on this figure, the various curves, except 
for the largest value of $E_{\rm PT2}$ of -0.2, almost coincide with the FCI curve at the scale
of the figure.


In table \ref{tab5}, the convergence of the dissociation energy, equilibrium distance, 
second-derivative, and non-parallelism error 
for the variational and full CIPSI energy curves as a function of $E_{\rm PT2}$ is reported. 
A first remark is that the spectroscopic values obtained 
from the variational and the full CIPSI curves at constant $E_{\rm PT2}$ are 
almost identical (in contrast with CIPSI 
at constant number of determinants, see Table \ref{tab2}). Of course, it is 
expected since the difference between the variational and full CIPSI 
energies is imposed to be constant. Note that the (very) small differences observed are 
because imposing a strict constant value of $E_{\rm PT2}$ is not possible 
due to the integer character of the number of determinants.
A second remark is that the convergence 
of the spectroscopic quantities as a function of the number of determinants 
(for each $E_{\rm PT2}$ an estimate of the average number of determinants along the energy curve 
is given in parentheses) is more rapid than when using CIPSI with a constant number 
of determinants, thus illustrating the efficiency of the constant $E_{\rm PT2}$ CIPSI approach.

\begin{table*}[t]
\begin{center}
\begin{tabular}{|l|c|c|c|c|}
\hline
\multicolumn{1}{|l}{} & \multicolumn{4}{c|}{CIPSI: Variational/full}\\
\hline
$E_{\rm PT2}$(a.u.) & $R_{eq}$ & $D_0$  & $k$ & $\Delta$ \\
\hline
-0.2 ($\sim$ 1.6 $10^2$ dets)  & 1.458/1.458   & 57.5/57.3 &  0.82/0.83 & 0.009/0.009\\
\hline
-0.1 ($\sim$ 5 $10^2$ dets)    &  1.450/1.451  & 50.0/51.0 & 0.83/0.82 & 0.009/0.009\\
\hline
-0.05 ($\sim$ 1.1 $10^3$ dets)  &  1.455/1.454  & 49.2/48.6 & 0.81/0.82 & 0.010/0.010\\
\hline
-0.02 ($\sim$ 3.5 $10^3$ dets)   &  1.459/1.459  & 44.6/45.5 & 0.78/0.78 & 0.011/0.011\\
\hline
-0.01 ($\sim$ 2 $10^4$ dets)  &  1.460/1.460  & 44.0/43.9 & 0.78/0.78 & 0.011/0.011\\
\hline
-0.008 ($\sim$ 3.5 $10^4$ dets) &  1.461/1.461  & 43.8/43.7 & 0.77/0.77 & 0.011/0.011\\
\hline
\hline
CIPSI $10^5$ dets$^a$ &  1.463/1.463  & 43.97/45.17 & 0.76/0.76 & 0.011/0.011\\
\hline
{\it i}-FCIQMC  &        & 45.00(11) &     &  \\
\hline
Exact NR       & 1.412  & 62.35 & 1.121 & 0.\\
\hline
\hline
\end{tabular}
\end{center}
\caption{Basis set =cc-pVDZ. Convergence of the spectroscopic 
quantities and non-parallelism error with CIPSI at constant $E_{\rm PT2}$ as 
function of $E_{\rm PT2}$.
For each $E_{\rm PT2}$, results obtained from the variational and full CIPSI energy curves
are given. Equilibrium distance in $\mathring{\rm A}$, dissociation energy in millihartree, 
and curvature in hartree/${\mathring{\rm A}}^2$.\\
$^a$ See Table \ref{tab2}
}
\label{tab5}
\end{table*}

\subsection{FN-DMC with CIPSI at constant $E_{\rm PT2}$}
\label{fndmcept2}

\begin{figure}[h]
\begin{center}
\includegraphics[width=0.9\columnwidth]{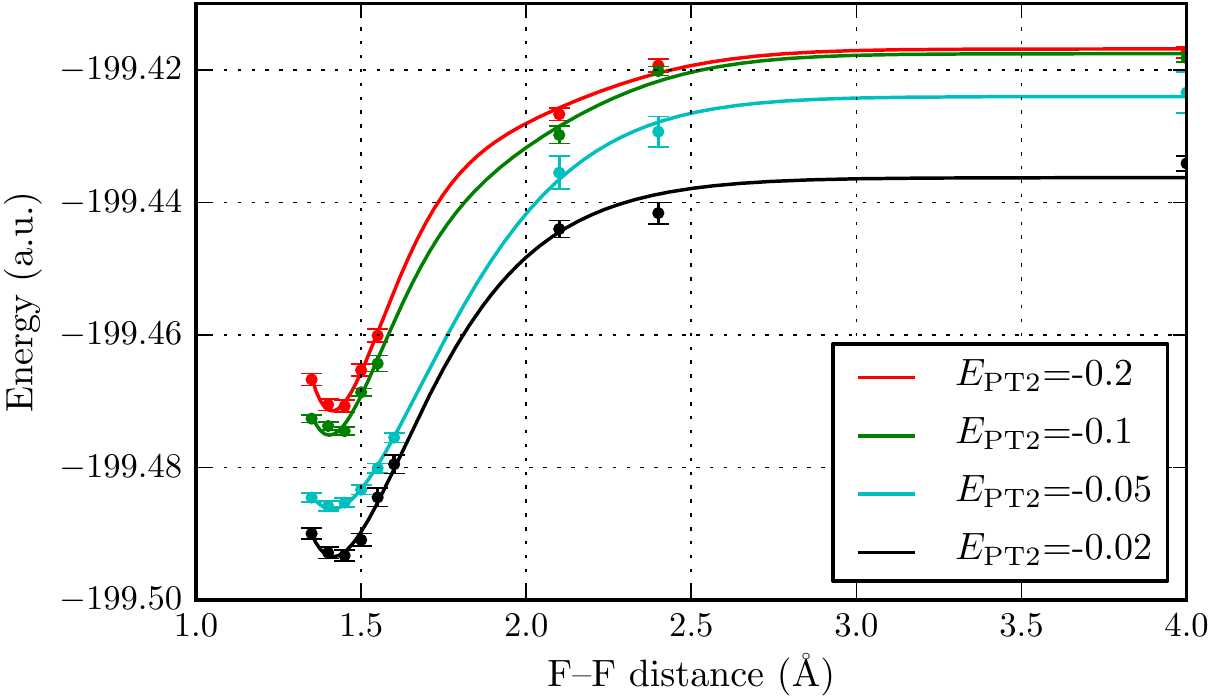}
\end{center}
\caption{FN-DMC energy curves for several values of $E_{\rm PT2}$ (VDZ basis set)}
\label{fig13}
\end{figure}

FN-DMC energy curves obtained from CIPSI-VDZ reference wavefunctions at constant $E_{\rm PT2}$
are presented in Fig.\ref{fig13}. Corresponding spectroscopic values and non-parallelism errors 
are reported in Table \ref{tab5}. To compare with, FN-DMC values using a constant number 
of 10$^4$ determinants are also given (taken from Table \ref{tab3}). FN-DMC spectroscopic quantities 
using trial wavefunctions at constant $E_{\rm PT2}$ systematically improve 
when decreasing $E_{\rm PT2}$ from -0.2 a.u. to -0.05 a.u. For this latter value, results are 
much improved with respect to FN-DMC values at constant number of determinants: 
The error in the dissociation energy is greatly reduced from 6.7 to 2.2 mEh, 
the equilibrium distance and curvature are found of comparable quality and, 
the non-parallelism error is significantly reduced to 0.0018, the best value obtained in this work. 
However, for the smaller value $E_{\rm PT2}$=-0.02 results deteriorate and 
become close to those obtained with FN-DMC with a fixed number of determinants. 
This result is of course expected since in the limit of a vanishing value for the second-order 
energy correction, the CIPSI algorithm at constant $E_{\rm PT2}$ reduces to the standard one.

\begin{figure}[h]
\begin{center}
\includegraphics[width=0.9\columnwidth]{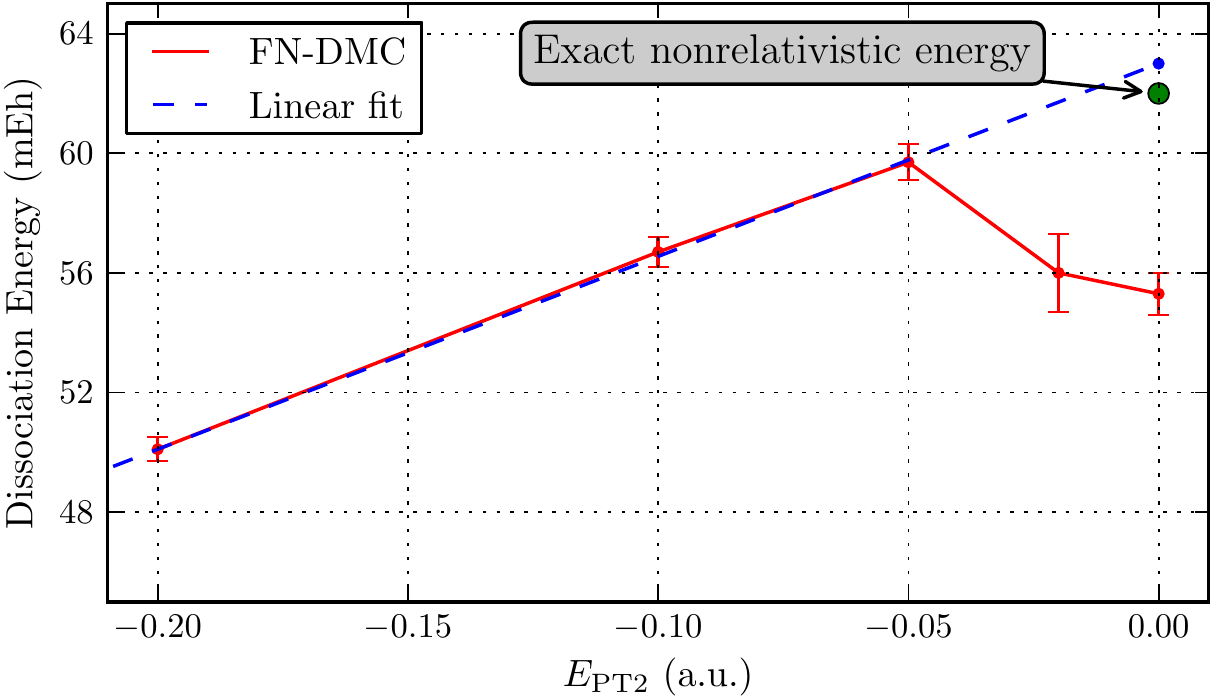}
\end{center}
\caption{FN-DMC/VDZ dissociation energies obtained for different values of $E_{\rm PT2}$}
\label{fig14}
\end{figure}

\begin{table}[h]
\begin{center}
\begin{tabular}{|l|c|c|c|c|}
\hline
$E_{\rm PT2}$(a.u.) &$R_{eq}$ & $D_0$ & $k$ & $\Delta$\\
\hline
\hline
-0.2 ($\sim$ 1.6 10$^2$ dets) &  1.442   & 50.1(4) & 0.839& 0.0045 \\
-0.1 ($\sim$ 5 10$^2$ dets) &  1.433   & 56.7(5)  & 1.190 & 0.0031\\
-0.05 ($\sim$ 1.1 10$^3$ dets) &  1.431   & 59.7(6)  & 1.131 & 0.0018\\
-0.02 ($\sim$ 3.5 10$^3$ dets) &  1.429   & 56.0(13) & 1.125 & 0.0028 \\
\hline
\hline
10$^4$ nodes                   & 1.428  &   55.3  & 1.117 & 0.0030\\
\hline
\hline
Exact       & 1.412  & 62.0 & 1.121 & 0. \\
\hline
\hline
\end{tabular}
\end{center}
\caption{Basis set=cc-pVDZ. FN-DMC spectroscopic values and non-parallelism errors using
	CIPSI-VDZ trial wavefunctions at constant $E_{\rm PT2}$. Equilibrium distance 
in $\mathring{\rm A}$, dissociation energy in millihartree,
and curvature in hartree/${\mathring{\rm A}}^2$.}
\label{tab5}
\end{table}
In Fig.\ref{fig14} the FN-DMC dissociation energies obtained for the different 
values of $E_{\rm PT2}$ are plotted. Two regimes are clearly observed. 
For $E_{\rm PT2}$=-0.2 a.u.,-0.1 a.u, and -0.05 a.u, the dissociation energy increases 
almost linearly as a function of the second-order correction (see, dashed line of the figure). 
Quite remarkably, when extrapolating the results (via a simple quadratic extrapolation)
to vanishing $E_{\rm PT2}$, a dissociation energy close to the exact value of 62.0 mEh is obtained. 
In the second regime corresponding to small values of $E_{\rm PT2}$ the curve 
reaches a maximum somewhere between -0.02 a.u. and -0.05 a.u. and, then, decreases down to 
a value at $E_{\rm PT2}$ =0 close to the FN-DMC result obtained above with the quasi-FCI trial wavefunction
(55.3 mEh). The existence of these two regimes 
is particularly striking and is interpreted here as follows. In the first regime corresponding to 
large values of $E_{\rm PT2}$, the determinants entering first the CIPSI expansion are those having 
a large coefficient in the exact wavefunction expansion. These determinants are typically 
associated with the multi-reference character of the system (static correlation contributions). 
In the second regime, many more determinants of much smaller weights enter the expansion. 
Their role is to build up dynamical correlation effects or, equivalently,
to better describe the small-distance details of the electron-electron interaction.
Determinants contributing to the first regime are expected to be weakly 
dependent on such small-distance details and, thus, on the quality of the basis set. 
This could be the reason why in the large-$E_{\rm PT2}$ regime the dissociation energy
extrapolates to the exact value independent on the basis set. 
In sharp contrast, at smaller $E_{\rm PT2}$ the numerous determinants with small weights that enter
in the expansion to describe the local details of the wavefunction are much more basis-set dependent. 
Thus, in this regime the behavior of the curve is expected to be strongly dependent on the specific
basis set employed.


\subsection{Graphical summary: F$_2$ curves at various levels of approximation}
\label{graphi}
\begin{figure}[h]
\begin{center}
\includegraphics[width=0.9\columnwidth]{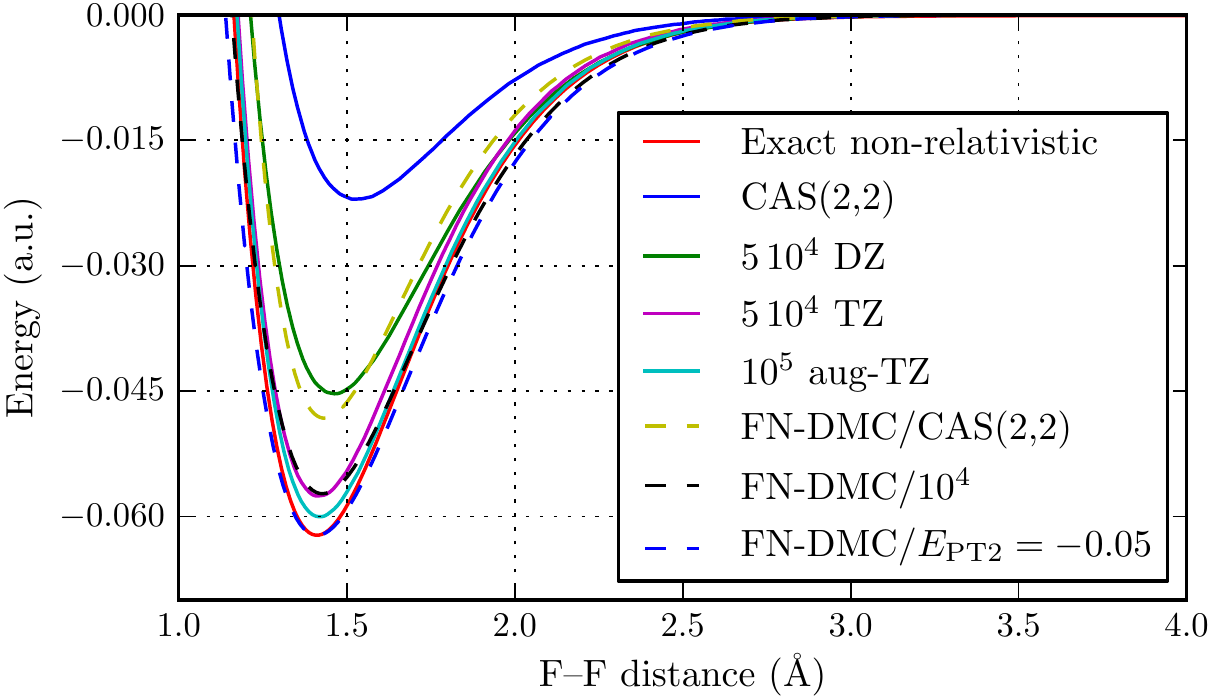}
\end{center}
\caption{F$_2$ potential energy curves using the various approaches presented in this work. 
Comparison with the estimated exact nonrelativistic curve. 
Each energy curve has been shifted down to impose at large distance the exact atomic limit.}
\label{fig15}
\end{figure}

In figure \ref{fig15} the main results of this work are summarized by showing on the same graph the 
various energy curves obtained. For the sake of clarity, each energy curve has been shifted down 
by the constant leading to a dissociation toward the exact (infinite basis) nonrelativistic atomic limit.
All curves are located between the poor-quality CASSCF curve (upper curve) 
and the estimated exact nonrelativistic fixed nucleus curve (lowest solid line). 
For the sake of clarity, energy curves obtained with deterministic calculations are represented by 
solid lines, while dashed lines are used for FN-DMC energy curves.
CIPSI energy curves using the VDZ, VTZ, and AVTZ basis sets are represented by the three 
curves (solid lines) between the two extreme CASSCF and exact curves.
As discussed above, these curves are almost converged in number of determinants and may be considered as 
a good approximation of the full CI limit in each case. Increasing the basis set 
has a clear important impact on the quality of the results. With the largest aug-VTZ basis set, the 
energy curve obtained is the best one, except for the FN-DMC curve obtained with a CIPSI-constant 
$E_{\rm PT2}$ 
trial wavefunction which is the most accurate energy curve calculate in this work. 
With the larger QZ basis set results continues to improve (not done here, our objective being 
not to increase indefinitely the basis set). The corresponding energy curve at a quasi FCI level is  
given in \cite{bytautas} and is found to be of a similar quality the our best FN-DMC-DZ curve.
The three FN-DMC curves are represented using a dashed line.
By increasing order of quality, they correspond to using i.) the CASSCF nodes ii) the 
CIPSI/VDZ nodes, and iii) the CIPSI/VDZ nodes at constant $E_{\rm PT2}$. Using CASSCF nodes the FN-DMC energy 
curve is much improved with respect to the variational curve obtained with the CASSCF wavefunction. 
Roughly speaking it is of quality of the quasi-FCI curve obtained with the VDZ basis set.
Now, taking CIPSI-VDZ nodes DMC results are improved to a quality typical of a FCI-VTZ calculation. 
Only when considering a nodal construction based on a variable number of determinants 
in a CIPSI-VDZ framework, DMC results surpass the quality of a FCI-AVTZ calculations. 
Let us emphasize that the preceding conclusions are essentially 
based on an overall quality as measured by the non-parallelism error. When having a closer look at 
the overall shape of the curves it is clear that the improvement for the two best curves are not 
uniform. When comparing the two best calculations, the near FCI-AVTZ and DMC/VDZ constant $E_{\rm PT2}$ 
curves, 
it is clear that the latest is much better near the equilibrium geometry. However, in the 
intermediate region, it is no longer true and the FCI calculation performs better thanks to its 
large number of determinants. The latter remark illustrates the fact that there is 
still some for improving the evolution of the nodes in this intermediate region.

\section{Summary}
\label{conclusion}
In this work we have investigated the quality of the F$_2$ potential energy curves calculated 
with Fixed-Node DMC using FCI-type expansions as trial wavefunctions.
Multi-determinantal CI wavefunctions have been constructed with the CIPSI method  
that selects iteratively the determinants contributing the most to the wavefunction,
as determined by first-order perturbation theory. A major advantage of CIPSI 
is to keep limited the expansion size since only those multiple-particle excitations that contribute 
the most in each sector of excitations (single, double, triple etc.) are considered. 
Quantitatively, to obtain CIPSI energy curves converged to FCI with an accuracy 
of the millihartee for the three basis sets used here (Dunning's cc-pVDZ, cc-pVTZ, and aug-cc-pVTZ) 
requires a number of determinants of at most several tens of thousands
(see, Figures \ref{fig1},\ref{fig2},\ref{fig3},\ref{fig4},\ref{fig5}, and \ref{fig6}).
It is remarkable that such convergence is possible with a number of determinants representing 
only a tiny fraction of the whole Hilbert space: About 10$^{-7}$, 10$^{-15}$, and 10$^{-18}$ 
for the VDZ, VTZ, and AVTZ basis sets, respectively. In practice, having compact wavefunctions 
is essential for DMC since the trial wavefunction and its derivatives need to be computed 
at each Monte Carlo step (potentially, billions of such steps may be performed). 

The Fixed-Node Diffusion Monte Carlo calculations have been performed using CIPSI wavefunctions directly as 
they come from the output of the CI program. No Jastrow factor has been employed 
(apart from the exact electron-nucleus cusp condition imposed at very small distances)
and no stochastic many-parameter optimization of the trial wavefunction in a preliminary VMC calculation 
has been performed. From a practical point of view, we emphasize that it is an important aspect 
since it greatly facilitates the implementation of fully automated QMC codes.
This is certainly an important prerequisite to allow large-scale diffusion of stochastic approaches 
beyond the limited community of QMC experts. Remarkably, in all cases considered 
the fixed-node error is found to systematically decrease when increasing the number 
of selected determinants (see, Fig.\ref{fig8}). The control of the fixed-node error is thus made simpler: 
When the convergence of the FN curve as a function of the number of determinants is approximately reached, 
a nodal structure not too far from the best one attainable within the given atomic basis set can be expected.
However, having in mind to treat large molecular systems, it is not realistic 
to rely on the systematic increase of the basis set to improve nodal structures.
We have thus proposed an alternative strategy based on the construction of coherent nodes along the PES
instead of very accurate nodes independently for each geometry. For a not too large basis set, it makes 
a major difference since the multi-reference character of the wavefunction known to change
considerably along the PES is much better taken into account and the overall quality of 
the energy curves is improved (smaller non-parallelism errors).
The main idea is to avoid using a common number of selected determinants
at each geometry leading to an unbalanced description of the PES but, instead, to 
consider determinantal expansions having a geometry-dependent length.
In practice, it is implemented by stopping the CIPSI selection process once the second-order 
estimate of the energy correction between the variational CIPSI and FCI energies 
has reached some target value, independent 
on the geometry. Using such trial wavefunctions, we have verified that improved FN-DMC energy curves 
are obtained, thus confirming that the nodal structure is better described along the PES.
However, more work is needed to investigate how such results generalize 
to more complex molecular systems.\\
\\
{\it Acknowledgments.}
We would like to acknowledge interesting discussions with Nathalie Guihery and Jean-Paul Malrieu (Toulouse). 
AS and MC thank the Agence Nationale pour la Recherche (ANR) for support through 
Grant No ANR 2011 BS08 004 01. This work has been made through generous computational support from CALMIP
(Toulouse) under the allocation 2013-0510, and GENCI under the allocation x2014081738.

\bibliography{cipsi}{}

\begin{thebibliography}{10}

\bibitem{lester}
B.~L. Hammond, W.A.~Lester Jr., and P.J. Reynolds.
\newblock {\em Monte Carlo Methods in Ab Initio Quantum Chemistry}, volume~1 of
  {\em Lecture and Course Notes in Chemistry}.
\newblock World Scientific, Singapore, 1994.
\newblock World Scientific Lecture and Course Notes in Chemistry Vol.1.

\bibitem{rmp}
W.M.C. Foulkes, L.~Litas, R.G. Needs, and G.~Rajagopal.
\newblock {\em Rev. Mod. Phys.}, 73:33, 2001.

\bibitem{lecuyer}
P.~L'Ecuyer.
\newblock {\em Math. of Comput.}, 68:261, 1999.

\bibitem{mosko}
K.E. Schmidt and J.W. Moskowitz.
\newblock {\em J. Chem. Phys.}, 93:4172, 1990.

\bibitem{flad}
H.J. Flad, M.~Caffarel, and A.~Savin.
\newblock {\em Quantum Monte Carlo calculations with multi-reference trial wave
  functions}.
\newblock Recent Advances in Quantum Monte Carlo Methods. World Scientific
  Publishing, 1997.

\bibitem{fili}
C.~Filippi and C.J. Umrigar.
\newblock {\em J. Chem. Phys.}, 105:213, 1996.

\bibitem{braida}
B.~Bra\i{i}da, J.~Toulouse, M.~Caffarel, and C.~J. Umrigar.
\newblock {\em J. Chem. Phys.}, 134:0184108, 2011.

\bibitem{goddard}
A.~G. Anderson and W.A.~Goddard III.
\newblock {\em J. Chem. Phys.}, 132:164110, 2010.

\bibitem{fili_vb}
F.~Fracchia, C.~Filippi, and C.~Amovilli.
\newblock {\em J. Chem. Theory Comput.}, 8:1943, 2012.

\bibitem{bouabca}
T.~Bouab\c{c}a, B.~Bra\^{\i}da, and M.~Caffarel.
\newblock {\em J. Chem. Phys.}, 133:044111, 2010.

\bibitem{qmcgrid}
A.~Monari, A.~Scemama, and M.~Caffarel.
\newblock Large-scale quantum monte carlo electronic structure calculations on
  the egee grid.
\newblock In Franco Davoli, Marcin Lawenda, Norbert Meyer, Roberto Pugliese,
  Jan Węglarz, and Sandro Zappatore, editors, {\em Remote Instrumentation for
  eScience and Related Aspects}, pages 195--207. Springer New York, 2012.
\newblock http://dx.doi.org/10.1007/978-1-4614-0508-5\_13.

\bibitem{mitas}
M.~Bajdich, L.~Mit\'a\v{s}, G.~Drobn\`y, and L.K. Wagner.
\newblock {\em Phys. Rev. Lett.}, 96:240402, 2006.

\bibitem{sorella}
M.~Casula, C.~Attaccalite, and S.~Sorella.
\newblock {\em J. Chem. Phys.}, 121:7110, 2004.

\bibitem{rios}
P.~Lopez Rios, A.~Ma, N.D. Drummond, M.D. Towler, and R.J. Needs.
\newblock {\em Phys. Rev. E}, 74:066701, 2006.

\bibitem{cyrus}
C.J. Umrigar, J.~Toulouse, C.~Filippi, S.~Sorella, and R.G. Hennig.
\newblock {\em Phys. Rev. Lett.}, 98:110201, 2007.

\bibitem{canadian}
E.~Giner, A.~Scemama, and M.~Caffarel.
\newblock {\em Can. J. Chem.}, 91:879, 2013.

\bibitem{cipsi1}
B.~Huron, P.~Rancurel, and J.P. Malrieu.
\newblock {\em J. Chem. Phys.}, 58:5745, 1973.

\bibitem{cipsi2}
S.~Evangelisti, J.P. Daudey, and J.P. Malrieu.
\newblock {\em Chem. Phys.}, 75:91, 1983.

\bibitem{preprint1}
T.~Applencourt, E.~Giner, A.~Scemama, and M.~Caffarel.
\newblock Accurate non-relativistic ground-state energies for 3d transtion
  metal atoms with fn-dmc.
\newblock {\em preprint}, 2014.

\bibitem{preprint2}
A.~Scemama, E.~Giner, M.~Caffarel, C.~Hureau, and P.~Faller.
\newblock Petascale quantum monte carlo for metal-free and copper-containing
  peptides: Critical role of the fixed-node approximation.
\newblock {\em preprint}, 2014.

\bibitem{towler}
M.D. Towler.
\newblock {\em Quantum Monte Carlo, or, how to solve the many-particle
  Schrödinger equation accurately whilst retaining favourable scaling with
  system size}.
\newblock Wiley, 2011.

\bibitem{ency}
M.~Caffarel.
\newblock {\em Quantum Monte Carlo Methods in Chemistry}.
\newblock Encyclopedia of Applied and Computational Mathematics. Springer,
  2012.

\bibitem{matsen}
F.A. Matsen.
\newblock {\em Adv. Quantum Chem.}, page~59, 1964.

\bibitem{sherman}
W.H. Press, B.P. Flannery, S.A. Teukolsky, and W.T. Vetterling.
\newblock {\em Numerical Recipes in C : The Art of Scientific Computing}.
\newblock Cambridge University Press, 1992.

\bibitem{bender}
C.~F. Bender and E.~R. Davidson.
\newblock {\em Phys. Rev.}, 183:23, 1969.

\bibitem{buenker1}
R.~J. Buenker and S.~D. Peyerimholf.
\newblock {\em Theor. Chim. Acta}, 35:33, 1974.

\bibitem{buenker2}
R.~J. Buenker and S.~D. Peyerimholf.
\newblock {\em Theor. Chim. Acta}, 39:217, 1975.

\bibitem{buenker3}
R.~J. Buenker, S.~D. Peyerimholf, and W.~Butscher.
\newblock {\em Mol. Phys.}, 35:771, 1978.

\bibitem{bruna}
P.~J. Bruna, D.~S. Peyerimholf, and R.~J. Buenker.
\newblock {\em Chem. Phys. Lett.}, 72:278, 1980.

\bibitem{buenker-book}
R.~J. Buenker, S.~D. Peyerimholf, and P.~J. Bruna.
\newblock {\em Computational Theoretical Organic Chemistry}.
\newblock Reidel, Dordrecht, 1981.

\bibitem{harrison}
R.~J. Harrison.
\newblock {\em J. Chem. Phys.}, 94:5021, 1991.

\bibitem{en1}
P.S. Epstein.
\newblock {\em Phys. Rev.}, 28:695, 1926.

\bibitem{en2}
R.K. Nesbet.
\newblock {\em Proc. Roy. Soc.}, A230:312, 1955.

\bibitem{mp}
C.~M{\o}ller and M.S. Plesset.
\newblock {\em Phys. Rev.}, 46:618, 1934.

\bibitem{dunning}
T.H. Dunning.
\newblock {\em J. Chem. Phys.}, 90:1007, 1989.

\bibitem{bytautas}
L.~Bytautas, T.~Nagata, M.S. Gordon, and K.~Ruedenberg.
\newblock {\em J. Chem. Phys.}, 127:164317, 2007.

\bibitem{booth2}
D.~Cleland, G.H. Booth, C.~Overy, and A.~Alavi.
\newblock {\em J. Chem. Theory Comput.}, 8:4138, 2012.

\bibitem{f2exact}
L.~Bytautas, N.~Matsunaga, T.~Nagata, M.~S. Gordon, and K.~Ruedenberg.
\newblock {\em J. Chem. Phys.}, 127:204301, 2007.

\bibitem{f2exactII}
L.~Bytautas and K.~Ruedenberg.
\newblock {\em J. Chem. Phys.}, 130:204101, 2009.

\bibitem{davidson_exact}
E.R. Davidson, S.A. Hagstrom, S.J. Chakravorty, V.~Meiser Umar, and C.~Froese
  Fischer.
\newblock {\em Phys. Rev. A}, 47:3649, 1993.

\bibitem{persico}
C.~Angeli and M.~Persico.
\newblock {\em Theor. Chem. Acc.}, 98:117, 1997.

\end{thebibliography}
\bibliographystyle{unsrt}
\end{document}